\documentclass[12pt,reprint,aps,groupedaddress,nofootinbib,prd,twocolumn]{revtex4-2}

\usepackage[english]{babel}
\usepackage[utf8]{inputenc}
\usepackage{amsmath}
\usepackage{mathbbol}
\usepackage{amssymb}
\usepackage{tabularx}
\usepackage[normalem]{ulem}
\usepackage{bbold}
\usepackage{graphicx,amsfonts}
\usepackage{epsfig}
\usepackage[colorlinks=true,
linkcolor=blue,
urlcolor=blue,
citecolor=blue]{hyperref}
\usepackage{bm,color}
\usepackage{mathrsfs}
\usepackage{enumerate}
\usepackage{amsthm}
\usepackage{bbm}
\usepackage{comment}
\usepackage{physics}
\usepackage{url}
\usepackage[normalem]{ulem}

\begin{document}
	
\title{Extreme-mass-ratio inspirals into rotating boson stars: \\nonintegrability, chaos, and transient resonances}
	
\author{Kyriakos Destounis, Federico Angeloni, Massimo Vaglio and Paolo Pani} 
\affiliation{Dipartimento di Fisica, Sapienza Università di Roma, Piazzale Aldo Moro 5, 00185, Roma, Italy}
\affiliation{INFN, Sezione di Roma, Piazzale Aldo Moro 2, 00185, Roma, Italy} 
	
\begin{abstract}
General relativity predicts that black holes are described by the Kerr metric, which has integrable geodesics. This property is crucial to produce accurate waveforms from extreme-mass-ratio inspirals. Astrophysical environments, modifications of gravity and new fundamental fields may lead to nonintegrable geodesics, inducing chaotic effects. 
We study geodesics around self-interacting rotating boson stars and find robust evidence of nonintegrability and chaos. We identify islands of stability around resonant orbits, where the orbital radial and polar oscillation frequency ratios, known as rotation numbers, remain constant throughout the island. These islands are generically present both in the exterior and the interior of compact boson stars. A monotonicity change of rotation curves takes place as orbits travel from the exterior to the interior of the star. Therefore, configurations with neutron-star-like compactness can support degenerate resonant islands. This anomaly is reported here for the first time and it is not present in black holes. Such configurations can also support extremely prolonged resonant islands that span from the exterior to the interior of the star and are shielded by thick chaotic layers. We adiabatically evolve inspirals using approximated post-Newtonian fluxes and find time-dependent plateaus in the rotation curves which are associated with island-crossing orbits. Crossings of external islands give rise to typical gravitational-wave glitches found in non-Kerr objects. Furthermore, when an inspiral is traversing an internal island that is surrounded by a thick chaotic layer, a new type of simultaneous multifrequency glitch occurs that may be detectable with space interferometers such as LISA, and can serve as evidence of an extreme-mass-ratio inspiral around a supermassive boson star. 
\end{abstract}
	
\maketitle
	
\section{Introduction}
	
Ground-based gravitational-wave~(GW) interferometers are about to start a new observation run and will continue detecting GW signals from the coalescence of compact binaries~\cite{LIGOScientific:2020ibl,LIGOScientific:2021qlt,LIGOScientific:2021djp}, and possibly from other GW sources in the years to come. Till date, almost a hundred of compact binary mergers have been reported. Despite the fact that the majority of events are well-understood as either black-hole (BH) or neutron-star~(NS) or mixed BH-NS mergers, some puzzling ``mass-gap'' events, such as GW190814~\cite{LIGOScientific:2020zkf} and GW190521~\cite{LIGOScientific:2020iuh,LIGOScientific:2020ufj}, challenge standard formation scenarios and have motivated  exotic alternatives (see, e.g.,~\cite{CalderonBustillo:2020fyi}). With the GW event catalog ever increasing, the possibility of detecting exotic compact objects~(ECOs) other than BHs and NSs is worth exploring~\cite{Cardoso:2019rvt,Maggio:2021ans}.

We are now convinced that an important amount of non-luminous exotic matter, known as dark matter~\cite{Bertone:2004pz,Clowe:2006eq,Bergstrom:2009ib}, is paramount in the formation~\cite{Primack:1997av} and amalgamation of galaxies~\cite{Corbelli:1999af}, as well as in determining the earliest, current, and future state of the universe~\cite{Kolb:1990vq,Primack:1997av,Carroll:2000fy,DelPopolo:2007dna,Arbey:2021gdg,Perivolaropoulos:2021jda}. This raises the possibility of new fundamental particles comprising the missing cosmological mass. With scalar fields predominantly used to model early universe physics, the case arises that such fields could form equilibrium condensates, held together by their own gravity, through a mechanism known as gravitational cooling~\cite{Seidel:1993zk,Sanchis-Gual:2019ljs}. Such prototypical class of ECOs has been dubbed boson stars~(BSs), and their conceptualization dates back to the late 1960s~\cite{Bonazzola:1966,Feinblum:1968,Kaup:1968,Ruffini:1969} and 1980s~\cite{Colpi:1986ye,vanderBij:1987gi,Frieman:1988ut,Frieman:1989bx,Ferrell:1989kz,Grasso:1990,Tkachev:1991ka,Liddle:1992fmk}.
	
At the fundamental level, BSs are the simplest localized configurations of a complex scalar field, governed by classical equations, thus even if they have not yet been observed in nature, they still can serve as models for compact objects ranging from particles to stars and less dense galactic halos. In all these cases, BSs are endowed with a balance between the dispersive nature of scalar matter and the gravitational pull holding them together (see~\cite{Jetzer:1991jr,Schunck:2003kk,Liebling:2023} for reviews on various types of BSs). BSs do not have an event horizon nor a singularity, are asymptotically flat, and may exhibit stable light rings, isolated ergoregions, and super-extremal spins, which can lead to new GW phenomenology~\cite{Liebling:2023,Kesden:2004qx,Macedo:2013jja,Macedo:2013qea,Macedo:2016wgh,Cardoso:2017cfl,Guo:2019sns}. As a consequence, they are considered among the best models of ECOs and a proxy to test the nature of compact objects in the extreme-gravity regime~\cite{Cardoso:2019rvt,Berti:2006qt,Pacilio:2020jza,Vaglio:2023lrd}. Another important motivation concerns the fact that BSs (and their real-scalar counterparts, known as oscillons~\cite{Seidel:1991zh}) could form in the early universe~\cite{Coleman:1985ki,Kusenko:1997si} from large overdensities, being thus (meta)stable relics of inflation~\cite{Lozanov:2019ylm,Aurrekoetxea:2023jwd} and compelling dark-matter candidates~\cite{Lozanov:2023aez,Franciolini:2023osw}.

Static BS configurations are linearly stable to perturbations~\cite{Gleiser:1988rq,Gleiser:1989,Kusmartsev:1990cr,Ryan:1997,Guzman:2004jw,Macedo:2013jja,Macedo:2016wgh} and can consistently form dynamically from diffuse initial states~\cite{Liddle:1992fmk,Levkov:2018kau,Veltmaat:2018dfz,Amin:2019ums,Arvanitaki:2019rax}, as well as from collisions and binary mergers~\cite{Bernal:2006ci,Palenzuela:2006wp,Palenzuela:2007dm,Choptuik:2009ww,Bezares:2017mzk,Palenzuela:2017kcg,Bezares:2018qwa,CalderonBustillo:2020fyi,Helfer:2021brt,Bezares:2022obu,Croft:2022bxq,Evstafyeva:2022bpr,Siemonsen:2023hko}. On the other hand, spinning BSs were found to be unstable toward a bar-mode instability~\cite{Sanchis-Gual:2019ljs,DiGiovanni:2020ror}, unless scalar self-interaction terms in the model are sufficiently strong~\cite{Siemonsen:2020hcg}. This instability is not present for rotating BSs made by a massive vector field~\cite{DiGiovanni:2020ror}.

Till date, a variety of bosonic potentials have been considered~\cite{Colpi:1986ye,vanderBij:1987gi,Schunck:1999zu,Friedberg:1986tq,Guerra:2019srj,Herdeiro:2017fhv,Collodel:2017biu,Boskovic:2021nfs,Collodel:2022jly,Pombo:2023xkw} each one providing a different relation between the BS maximum mass and the underlying field-theory parameters. As a rule of thumb, strong self-interactions can make the maximum mass parametrically larger than in the free-scalar case, motivating alternative models for supermassive objects in galactic cores, which can mimic the shadows~\cite{Vincent:2015xta,Cunha:2015yba,Cunha:2016bjh,Cunha:2018acu,Herdeiro:2021lwl,Rosa:2022tfv} and particle orbits around ordinary BHs~\cite{Torres:2000dw,Macedo:2013jja}. 

Hence, it is of utmost importance to devise further tests in order to distinguish BSs through orbital dynamics and GW observations. In this work we focus on extreme-mass-ratio inspirals (EMRIs), consisting of a primary supermassive compact object (which we assume to be a self-interacting, compact, spinning BS) and a stellar-mass compact secondary. So far, geodesic and inspiral studies have been carried out at the equatorial plane of non-rotating~\cite{Kesden:2004qx,Diemer:2013zms,Macedo:2013jja,Macedo:2013qea,Brihaye:2014gua,Pombo:2023ody} and rotating BSs~\cite{Grandclement:2014msa,Grould:2017rzz,Zhang:2021xhp} (see also the recent~\cite{Delgado:2023wnj} for circular, equatorial EMRIs around hairy BHs~\cite{Herdeiro:2014goa} interpolating between a Kerr BH and a BS without self-interactions). The emitted GWs during circular equatorial geodesics and EMRIs were also analyzed in some BS models~\cite{Kesden:2004qx,Macedo:2013jja,Collodel:2021jwi,Zhang:2021ojz,Delgado:2023wnj}. 
However, a geodesic and inspiral analysis is lacking for generic, non-equatorial and non-circular, trajectories around spinning BSs. Generic orbits will not only help to further constrain the existence of supermassive BSs through their considerably more intricate waveform signal, but are also tools to examine the integrability of the underlying geodesic equations that govern particle motion. If geodesics around BSs are integrable, then the evolution is regular and no chaotic phenomena are present, as in the Kerr case. In this case, geodesics and EMRIs between rotating BSs and Kerr BHs will differ only by their different multipolar structure~\cite{Ryan:1995wh,Ryan:1997hg,Ryan:1997kh,Pacilio:2020jza,Vaglio:2022flq,Vaglio:2023lrd}. Alternatively, if geodesics around rotating BSs break integrability then direct and indirect chaotic imprints emerge that are clearly distinguishable both at the orbital~\cite{Apostolatos:2009vu,Lukes-Gerakopoulos:2010ipp,Contopoulos:2011dz,Lukes-Gerakopoulos:2012qpc,Zelenka:2017aqn,Zelenka:2019nyp,Lukes-Gerakopoulos:2021ybx,Destounis:2020kss,Deich:2022vna,Chen:2022znf,Destounis:2023cim,Pan:2023wau} and GW level~\cite{Destounis:2021mqv,Destounis:2021rko,Destounis:2023gpw}. 
	
It is the main goal of this work to study generic orbits and EMRIs around supermassive self-interacting BSs in order to assess if their exotic multipolar structure~\cite{Ryan:1995wh,Ryan:1997hg,Ryan:1997kh,Pacilio:2020jza,Vaglio:2022flq,Vaglio:2023lrd} breaks integrability and gives rise to imprints of chaos. If geodesics around such objects are nonintegrable, then GW observations from future spaceborne detectors like the Laser Interferometer Space Antenna (LISA)~\cite{Glampedakis:2005hs,LISA:2017pwj,Baibhav:2019rsa,Amaro-Seoane:2022rxf,LISA:2022kgy,Karnesis:2022vdp}, TianQin~\cite{TianQin:2015yph} and Taiji~\cite{Ruan:2018tsw,Ruan:2020smc} may lead to distinguishable effects that can break the degeneracy between supermassive BHs and BSs in galactic centers\footnote{Note that also EMRIs around stellar-mass compact objects would be potentially detectable by third-generation detectors such as the Einstein Telescope if the secondary is a subsolar compact object~\cite{Barsanti:2021ydd}.}. In what follows, we adopt geometrized units so that $G=c=1$.
	
\section{Geodesics and Chaos}
	
Generic stationary and axisymmetric spacetimes can be written as
\begin{equation}\label{line_element}	ds^2=g_{tt}dt^2+2g_{t\varphi}dtd\varphi+g_{rr}dr^2+g_{\theta\theta}d\theta^2+g_{\varphi\varphi} d\varphi^2,
\end{equation}
where the metric tensor components are, in general, functions of $r$ and $\theta$, and the coordinate system $\left(t,r,\theta,\varphi\right)$ can be, e.g., of Boyer-Lindquist type~\cite{Boyer:1966qh} or quasi isotropic~\cite{Misner:1973prb}. The motion of a secondary point-particle orbiting the spacetime geometry of the primary compact object (as defined by Eq.~\eqref{line_element}) is described by the geodesic equations,
\begin{equation}\label{geodesic_equation}
\ddot{x}^\kappa+\Gamma^\kappa_{\lambda\nu}\dot{x}^\lambda\dot{x}^\nu=0,
\end{equation}
where $\Gamma^\kappa_{\lambda\nu}$ are the Christoffel symbols of spacetime, $x^\kappa$ is the four-position vector of the orbit, and the overdots denote differentiation with respect to proper time $\tau$. 
The geodesic equation~\eqref{geodesic_equation} breaks down into four equations of motion, one for each coordinate component of the particle's trajectory, $x^\kappa(\tau)$. This system of second order, coupled differential equations can be considerably simplified using spacetime symmetries. The spacetime~\eqref{line_element} assumed in this work possesses two Killing vector fields resulting from stationarity and axisymmetry (the metric tensor is $t$- and $\varphi$-independent), yielding two conserved quantities throughout geodesic motion, i.e. the specific energy and $z$-component of the angular momentum of the particle,
\begin{equation}\label{energy_momentum}
-E/m=g_{tt}\dot{t}+g_{t\varphi}\dot{\varphi},\,\,\,\,\,\,\,\,\,
L_z/m=g_{t\varphi}\dot{t}+g_{\varphi\varphi}\dot{\varphi},
\end{equation}
where $m$ is the mass of the orbiting particle. Rearranging Eq.~\eqref{energy_momentum} we obtain two first-order, decoupled differential equations for the $t$ and $\varphi$ momenta as
\begin{equation}\label{tphi_dot}
\dot{t}=\frac{E g_{\varphi\varphi}+L_zg_{t\varphi}}{m\left(g^2_{t\varphi}-g_{tt}g_{\varphi\varphi}\right)},\,\,\,\,\,\,\,\,\,
\dot{\varphi}=\frac{E g_{t\varphi}+L_zg_{tt}}{m\left(g_{tt}g_{\varphi\varphi}-g^2_{t\varphi}\right)}\,,
\end{equation}
which can be solved once $r(\tau)$ and $\theta(\tau)$ are known.
The two remaining equations of motion for $r$ and $\theta$ are, in general, coupled and of second differential order. Test particles in geodesic motion provide a third constant of motion, namely the conservation of their rest mass (or equivalently their four-velocity $g_{\lambda\nu}\dot{x}^\lambda\dot{x}^\nu=-1$) which leads to a constraint equation of the form
\begin{equation}\label{constraint_equation}
\dot{r}^2+\frac{g_{\theta\theta}}{g_{rr}}\dot{\theta}^2+V_\text{eff}=0,
\end{equation}
with $V_\text{eff}$ being a Newtonian-like potential,
\begin{equation}
V_\text{eff}\equiv\frac{1}{g_{rr}}\left(1+\frac{g_{\varphi\varphi} E^2+g_{tt}L_z^2+2g_{t\varphi}E L_z}{m^2\left(g_{tt}g_{\varphi\varphi}-g_{t\varphi}^2\right)}\right),
\end{equation}
that characterizes bound geodesic motion. When $V_\text{eff}=0$ the resulting curve is called curve of zero velocity~(CZV) since $\dot{r}=\dot{\theta}=0$ there. 

If a hypothetical rank-two (or higher-rank) Killing tensor field exists, then the motion of $r(\tau)$ and $\theta(\tau)$ decouples into first-order differential equations. A special case where a rank-two Killing tensor exists is the Kerr solution where the role of the separation constant is played by the famous Carter constant~\cite{Carter:1968rr}. However, since the compact object we will assume throughout our analysis has a different multipolar structure than that of a Kerr BH (and furthermore it is known only numerically), it is unlikely that a separation, Carter-like, constant (or any other higher-rank Killing tensor) exists. 
Thus, to evolve orbits we will use the coupled second-order differential equation system for $r$ and $\theta$, together with Eqs.~\eqref{tphi_dot} and~\eqref{constraint_equation}, without any further symmetry assumptions besides stationarity and axisymmetry.
	
The existence of Carter's separation constant is not only a useful tool to evolve trajectories in Kerr spacetime faster, but rather implies an important aspect of geodesics, namely their integrability. Kerr spacetimes have four degrees of freedom. Stationarity and axisymmetry leads to the reduction of degrees of freedom to two. Taking into account the existence of the Carter constant and the conservation of the rest mass of the test particle reduces the degrees of freedom of orbital motion to zero; geodesics around Kerr BHs are integrable and do not present chaotic features~\cite{Contopoulos_book,Apostolatos:2009vu,Lukes-Gerakopoulos:2010ipp,Lukes-Gerakopoulos:2012qpc,Destounis:2020kss,Destounis:2021mqv,Destounis:2021rko,Lukes-Gerakopoulos:2021ybx,Destounis:2023gpw}. Unfortunately, Carter's symmetry is extremely fragile and in many occasions is broken by simply deforming the multipolar structure of spacetime by considering accretion disks and BH environments~\cite{Zipoy:1966,Voorhees:1970ywo,Cardoso:2021wlq,Cardoso:2022whc,Destounis:2022obl,Polcar:2022bwv,Figueiredo:2023gas}, modifications of gravity~\cite{Destounis:2020kss,Deich:2022vna}, neutron stars~\cite{Hartle:1968,Manko:2000ud,Pappas:2012nv,Pappas:2016sye}, exotic compact objects~\cite{Manko_1992,Manko:2000sg}, or in a parameterized way by introducing agnostic deformations to Kerr; a class of metrics called bumpy or non-Kerr BHs~\cite{Collins:2004ex,Glampedakis:2005cf,Vigeland:2009pr,Johannsen:2011dh,Vigeland:2011ji,Johannsen:2012mu,Johannsen:2013szh,Emparan:2014,Cardoso:2014rha,Rezzolla:2014mua,Konoplya:2016jvv,Moore:2017lxy}. In these cases, the integrability property may be broken, leading to chaotic effects~\cite{Contopoulos_book}. Full-blown, ergodic chaos is not expected to occur in astrophysical scenarios, such as EMRIs, but nonintegrability-like effects are anticipated even when geodesics are integrable~\cite{Flanagan:2010cd,Berry:2016bit,Speri:2021psr,Gupta:2022fbe} due to dissipation. This is mainly due to the manifestation occurring around transient orbital resonances which are expected to affect EMRI evolution and parameter estimation, though nonintegrable EMRIs are even more likely to amplify these effects and can introduce clear chaotic phenomena~\cite{Apostolatos:2009vu,Lukes-Gerakopoulos:2010ipp,Contopoulos:2011dz,Lukes-Gerakopoulos:2012qpc,Lukes-Gerakopoulos:2012qpc,Zelenka:2017aqn,Zelenka:2019nyp,Destounis:2021mqv,Destounis:2021rko,Lukes-Gerakopoulos:2021ybx,Destounis:2023gpw}.	
A generic orbit of an integrable system can be described by its revolution frequency $\omega_\varphi$ and two libration-like frequencies; the frequency of oscillation from the periapsis to the apoapsis and back, $\omega_r$, and the oscillation frequency through the equatorial plane, $\omega_\theta$. Generic orbits possess irrational ratios of the above frequencies and these orbits fill densely the available phase space of a three-dimensional torus. On the other hand, resonant (periodic) orbits, have commensurate (rational) ratios of orbital frequencies which means that the particular orbits are returning to their initial position after some revolutions depending on their periodicity, thus they are not phase-space filling. When orbital resonances are encountered during inspirals of equal mass binaries, e.g. in the kHz band of LIGO/Virgo/KAGRA detectors, they do not affect the evolution since the inspiral is extremely rapid. However, when the binary is highly asymmetric, i.e. an EMRI, then the adiabatic nature of the secondary's motion can experience orbital resonances for a significant number of cycles and lead to substantial effects, such as cumulative dephasing and putative erroneous parameter estimation~\cite{Flanagan:2010cd,Berry:2016bit,Speri:2021psr}. 
	
When a nonintegrable perturbation is introduced to the system, two theorems, namely the Kolmogorov-Arnold-Moser (KAM)~\cite{Moser:430015,Arnold_1963} and Poincar\'e-Birkhoff~\cite{Birkhoff:1913} theorems, dictate how the phase space structure is altered around resonant points. KAM theorem ensures that when the orbits are sufficiently away from resonances, the system behaves as if it were integrable. The trajectories in the phase space lie on a torus defined by the integrals of motion and successive intersections of the orbits on a perpendicular 2-dimensional surface (the Poincaré surface of section) form curves that organize around a common, fixed, central point. These are called KAM curves and the whole structure is known as Poincar\'e map, whose central point corresponds to a planar circular orbit. Close to periodic orbits, the KAM curves disintegrate into two sets of periodic points in the Poincar\'e map, in accord to Poincar\'e-Birkhoff theorem; the stable ones which are surrounded by islands of stability (resonant islands) and the unstable ones where chaotic orbits emanate and surround the islands of stability with thin layers. The whole structure around resonances of nonintegrable systems is called a Birkhoff chain (see Fig. 2 in~\cite{Lukes-Gerakopoulos:2010ipp} for an illustration). The crucial aspect of resonant islands, and their significance in EMRIs, is the fact that the rational ratio of the orbital frequencies $\omega_r/\omega_\theta$ is shared throughout the island for all geodesics that occupy it. Put in other words, integrable EMRIs experience resonances that occupy a zero-volume point in phase space while nonintegrable EMRIs exhibit prolonged resonances where the secondary is locked in perfect resonance for a rather significant amount of revolutions (order of few hundreds of cycles, for typical mass ratios and depending on the non-Kerr object~\cite{Destounis:2021mqv,Destounis:2021rko,Lukes-Gerakopoulos:2010ipp,Lukes-Gerakopoulos:2021ybx,Destounis:2023gpw}) without taking into account pre- and post-resonant effects~\cite{Ruangsri:2013hra} and conservative effects from gravitational self-force~\cite{Barack:2009ux} which are essential for integrable EMRIs to show signs of transient resonant phenomena.
	
The existence of ergodic chaos and islands of stability around periodic points are direct and indirect signatures of nonintegrability, respectively, therefore sketching a detailed Poincar\'e map can unravel if the system under study is chaotic. When the perturbation introduces a slight nonintegrable deformation to the system there are other techniques to assess if and at which initial conditions of geodesics the aforementioned phenomena manifest themselves. The rotation number is one of the most helpful tool to find regions of interest in order to search for islands of stability. We calculate it by tracking the angle $\vartheta$ between successive intersections on KAM curves, relative to the fixed central point of the Poincar\'e map. The rotation number is defined as the summation of all angles $\vartheta$ measured between consecutive intersections, i.e.~\cite{Contopoulos_book}
\begin{equation}\label{rotation}
\nu_\vartheta=\frac{1}{2 \pi N}\sum_{i=1}^{N}\vartheta_i,
\end{equation}
where $N$ is the number of angles measured. When $N\rightarrow\infty$, Eq.~\eqref{rotation} asymptotes to the orbital frequency ratio $\nu_\vartheta=\omega_r/\omega_\theta$. Calculating consecutive rotation numbers for different initial conditions of orbits, by smoothly varying one of the parameters of the system while keeping the rest fixed, leads to a rotation curve.
	
Integrable systems show monotonous rotation curves, while nonintegrable systems display discontinuities in the monotonicity through the formation of transient plateaus with a non-zero width when geodesics transverse resonant islands. Inflection points can also appear when trajectories pass through unstable periodic points. Nevertheless, by changing the initial conditions properly, the orbit can be driven through the island and give rise to a plateau. So far, all studies have dealt with compact objects that are either non-Kerr or bumpy in nature and either possess an event horizon or have serious causal structure pathologies. In the following sections, we will examine the characteristic features of rotating self-interacting BSs with geodesics and approximate EMRI evolutions in order to first assess whether geodesic motion in such spacetimes is nonintegrable and, then, to establish various phenomenological imprints of chaos in the associated GW signal.

Before we proceed, we note that all numerical evolution have been performed with respect to the intertial time of the detector at infinity and not with respect to proper time in order to extract the corresponding GW emission of EMRIs. To achieve that, we have transformed the equations of motion from proper to inertial time by the use of the chain rule for first and second derivatives with respect to proper time, e.g. $\dot{r}=dr/d\tau=(dr/dt)(dt/d\tau)=\dot{t}\, r^\prime $, where we have defined $r^\prime\equiv dr/dt=\dot{r}/\dot{t}$. Equivalent equations hold for $\theta$, $\varphi$ and $t$ as well as for the constraint equation~\eqref{constraint_equation}.
	
\section{Self-interacting rotating BSs}

The equilibrium configurations of rotating BSs can be constructed starting from the globally $U(1)$-invariant action
\begin{equation}
{\cal S}=\int d^4 x\sqrt{-g}
\left[\frac{R}{16\pi}-{\cal L}_\phi\right]\ ,
\end{equation} 
which describes the dynamics of a complex, massive scalar field $\phi$, minimally coupled to gravity. The Lagrangian $\mathcal{L}_\phi$  considered in this work is characterized by a mass parameter $\mu$, plus quartic repulsive corrections, controlled by the coupling $\lambda$, which add linearly to the kinetic term as
\begin{equation}
{\cal L}_\phi=-\frac{1}{2}g^{\alpha\beta}\phi^*_{,\alpha}\phi_{,\beta}-\frac{1}{2}\mu^2|\phi|^2-\frac{1}{4}\lambda|\phi|^4 \ . \label{lagrangian}
\end{equation}
The requirement of stationarity and axisymmetry, leads to the following ansatz for the scalar field
\begin{equation}
\phi=\phi_0(r,\theta)e^{i(s\varphi-\Omega t)}\ ,
\label{eq:field_ansatz}
\end{equation} 
where $\Omega>0$ is the field's angular frequency, which determines the phase evolution in time, while $s$ is an integer called the azimuthal (or rotational) winding number and can be proven to correspond to the ratio between the conserved angular momentum and particle number~\cite{Schunck1996,Yoshida:1997qf,Schunck:2003kk}. The ansatz~\eqref{eq:field_ansatz} ensures that the stress-energy tensor $T_{\alpha\beta}[g^{\alpha\beta},\phi,\partial_\alpha \phi]$ does not depend on $t$ and $\varphi$ and sources a spacetime metric with time and azimuthal symmetry.

In this work, following~\cite{Ryan:1995wh,Vaglio:2022flq}, we consider spinning BSs with large self-interactions, i.e. characterized by $\lambda/\mu^2\gg1$. In this limit the $(r,\theta)$ scalar profile is approximately constant in the star's interior and one can neglect the radial and polar derivatives of the field  ($\partial_r\phi\sim0$, $\partial_\theta \phi \sim 0$), while assuming $\phi \sim 0$ in the exterior. This allows expressing the stress-energy tensor of the BS as that of a perfect fluid and to define the radius as in ordinary stars\footnote{In the general case, the stress-energy tensor of a BS contains anisotropic terms and the scalar field, although exponentially decreasing, extends up to infinity, so that the radius is conventionally defined as the value of the radial coordinate enclosing a sufficiently large fraction (typically $99\%$) of the total mass ~\cite{Kaup:1968,Ruffini:1969,Amaro-Seoane:2010pks}. In our case, instead, the scalar field has compact support and the radius is uniquely defined.}.  Furthermore, with an appropriate redefinition of the variables and parameters, the coupling constants in the Lagrangian can be factored out, so that each numerical solution corresponds to a one-parameter family of configurations, controlled by the effective mass parameter $M_B=\sqrt{\lambda}/\mu^2$~\cite{Vaglio:2022flq}. The dimensionful physical quantities characterizing each BS, such as its mass, radius and energy-density, can be obtained from the dimensionless ones, characterizing each numerical solution, by multiplying them by the required power of $M_B$ to match their dimension in mass and scale correspondingly as it changes. 
	
Adopting quasi-isotropic coordinates, the line element~\eqref{line_element} can be described through four independent functions $(\rho,\gamma,\omega,\alpha)$ of $(r,\theta)$ as
\begin{subequations}
\begin{align}
\nonumber \\[-10pt]
g_{tt}(r,\theta)&=-e^{\gamma(r,\theta)+\rho(r,\theta)}  \nonumber \\
& +\omega(r,\theta)^2 e^{\gamma(r,\theta)-\rho(r,\theta)} r^2\sin^2\theta\,, \\[10pt]
g_{rr}(r,\theta)&=\frac{g_{\theta\theta}(r,\theta)}{r^2}=e^{2\alpha(r,\theta)}\,, \\[10pt]
g_{t\varphi}(r,\theta)&=-\omega(r,\theta) e^{\gamma(r,\theta)-\rho(r,\theta)} r^2 \sin^2\theta\,, \\[10pt]
g_{\varphi\varphi}(r,\theta)&=e^{\gamma(r,\theta)-\rho(r,\theta)} r^2 \sin^2\theta\,. \\[-10pt]\nonumber
\end{align}
\end{subequations}
We constructed the solutions numerically following the method described in~\cite{Ryan:1997,Vaglio:2022flq} which makes use of an integral representation of the Einstein equations to set up an iterative integration scheme. The algorithm starts with a solution corresponding to a non-rotating BS defined on a two-dimensional grid of $r \in (10^{-6},10)M_B$ (which typically corresponds to $r \in (10^{-5},100)M$ in terms of the BS mass $M$), $\theta \in (0,\pi/2)$, and converges to a spinning configuration, close to the initial one, within roughly $150$ iterations, with a relative error of $\mathcal{O}(10^{-4})\%$ for all metric functions and the scalar profile. The details of the numerical implementation are provided in~\cite{Vaglio:2022flq}. 
	
The maximum mass that can be reached by these stars, scales as~\cite{Colpi:1986ye,Vaglio:2022flq}
\begin{equation}
M_{\rm max}\sim \gamma(\chi)M_B=\gamma(\chi)\frac{\sqrt{\lambda \hbar}}{m_s^2}M_p^3\ ,\label{eq_maxmass}
\end{equation}
where $m_s=\mu\hbar$ is the mass of the boson, $M_p$ the Planck mass and $\gamma(\chi)$ is a $\mathcal{O}(0.1)$ factor which depends on the dimensionless spin $\chi=J/M^2$ (where $J$ is the angular momentum of the solution). This means, for instance, that, for $\lambda \hbar \sim \mathcal{O}(10^{-80})$ and $m_s$ in the range $10^{-15}$--$10^{-12}$ eV, the model allows for compact stellar configurations with $M_{\rm max}$ in the range $10$--$10^7~M_\odot$. For $\lambda$ as large as $\lambda \sim \mathcal{O}(\hbar^{-1})$ the same range of $M_{\rm max}$ corresponds to $m_s \in (0.1,100)$ MeV.

In what follows, we consider rapidly-rotating supermassive configurations with $\chi\sim0.8$, same mass and decreasing compactness, whose properties are listed in Table~\ref{tab:configurations}.
All the configurations have topological genus 1~\cite{Vaglio:2022flq}, at variance with their vector counterparts, spinning Proca stars~\cite{Brito:2015pxa}, which have instead a spherical topology even when spinning.
\begin{table}[thp]
\begin{tabularx}{\linewidth}{
>{\hsize=0.75\hsize\linewidth=\hsize}X
>{\hsize=0.75\hsize\linewidth=\hsize}X
>{\hsize=0.75\hsize\linewidth=\hsize}X
>{\hsize=0.75\hsize\linewidth=\hsize}X
>{\hsize=0.75\hsize\linewidth=\hsize}X
>{\hsize=0.75\hsize\linewidth=\hsize}X}
\hline\hline
 & $\hspace{-0.21cm} M/M_\odot$ & $\,\,\chi$ & $\,\,\,\,\mathcal{C}$ & $\,\,\,\Omega$ &$M_B/M_\odot$\\\hline
\text{Case~1} & $10^6 $ & $0.8$ & $0.25$ & $0.73$ & $8.5 \times 10^6 $\\
\text{Case~2} & $10^6 $ & $0.8$ & $0.19$ & $0.80$& $9.0\times10^6 $\\
\text{Case~3} & $10^6$ & $0.8$ & $0.16$ & $0.82$& $1.0 \times 10^7$\\
\hline\hline
\end{tabularx}
\caption{Self-interacting, rotating BSs considered in this work. The three configurations have the same dimensionless spin and mass but different compactnesses and frequencies (due to the different effective coupling $M_B$). 
The compactness is defined by ${\cal C}\equiv M/R$, where $R$ is the star's perimeteral radius~\cite{Herdeiro:2015gia}.
The order of magnitude chosen for $M_B$ corresponds, considering $\lambda/\mu^2\sim \mathcal{O}(100) \gg 1$, to a mass of the boson $m_s=\hbar \mu \sim 10^{-14}\,{\rm eV}$.
}\label{tab:configurations}
\end{table}

\begin{figure}[htp]
\includegraphics[width=0.5\textwidth]{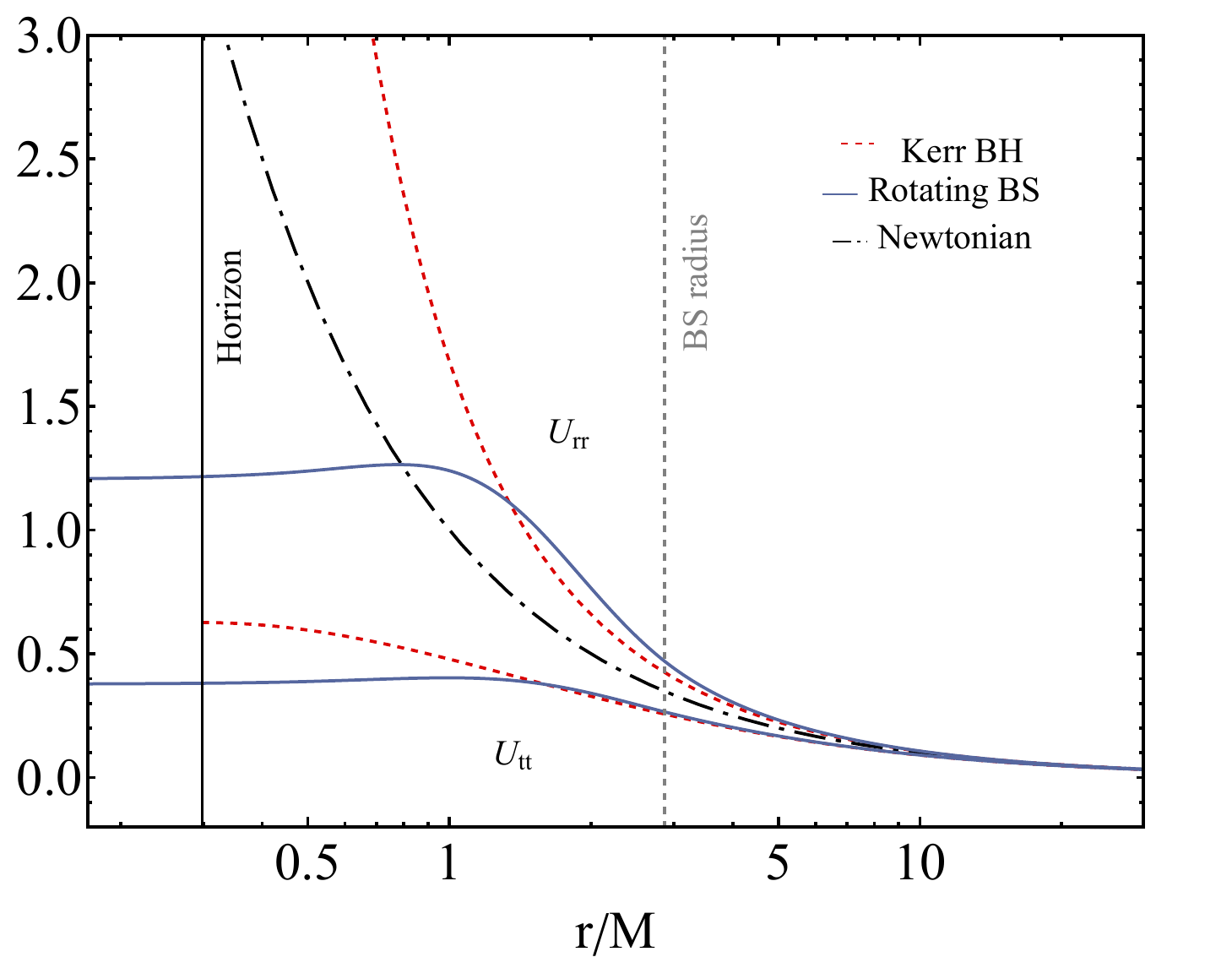}
\caption{Comparison between the Newtonian potential $M/r$ (black dashed) and the equatorial radial profiles of  $U_{tt}\equiv(g_{tt}+1)/2$ and $U_{rr}\equiv(g_{rr}-1)/2$ for the Kerr metric (red dashed), and the rotating BS (blue solid) corresponding to Case~1. The black solid and gray dashed vertical lines correspond to the location of the Kerr horizon and the BS radius in the quasi-isotropic radial variable $r/M$.}
\label{fig:potential}
\end{figure}

The configuration of Case~1 is the most compact one and corresponds to the maximum-mass solution for $\chi= 0.8$. A comparison with the Kerr solution, sharing the same mass and spin, is shown in Fig.~\ref{fig:potential}, in terms of the functions $U_{tt}\equiv(g_{tt}+1)/2$ and $U_{rr}\equiv(g_{rr}-1)/2$ evaluated on the equatorial plane. The corresponding Kerr event horizon and BS radius are also shown, as well as the Newtonian gravitational potential $M/r$ for an object of the same mass. 
 
All the configurations in Table~\ref{tab:configurations} are in a linearly perturbative stable branch of the mass-frequency diagram. It is known that BSs with no or weak self-interactions are subject to a dynamical non-axisymmetric instability that develops on short timescales~\cite{Sanchis-Gual:2019ljs}. Such instability has been shown to be quenched if the scalar self-interactions are sufficiently strong~\cite{Siemonsen:2020hcg}. This is precisely the limit in which our solutions are obtained, with the explicit value of $\lambda/\mu^2$ depending on the individual choices for $\lambda$ and $\mu$, once $M_B$ is fixed. Rotating BSs in this regime have also recently been formed dynamically in numerical simulations as a result of a binary coalescence, starting from non-rotating components in a quasi-circular orbit~\cite{Siemonsen:2023hko}. 
	
\begin{figure}[htp]
\hspace*{-0.3cm}
\includegraphics[width=0.5\textwidth]{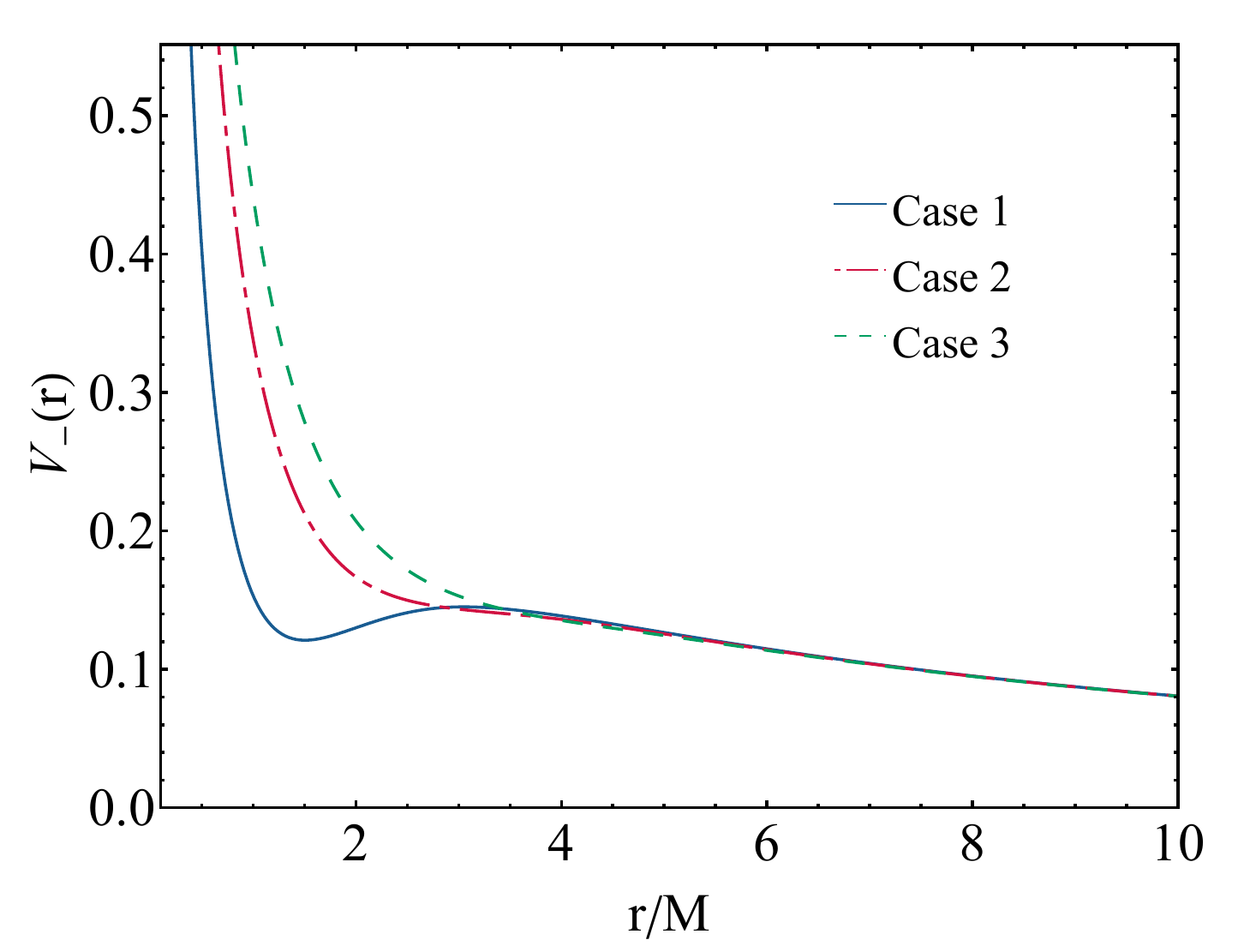}
\caption{Equatorial effective potentials $V_{-}(r)$ for the BS configurations corresponding to our three case studies. The BS in Case~1 has two light rings, corresponding to the stationary points $V'_{-}(r)=0$, one of which is unstable and outside the star, while the other is stable and inside it. Stationary points of $V_{-}(r)$ correspond to counter-rotating circular photon orbits, while $V_{+}$ (not shown here) is associated to co-rotating photon orbits which are not present for these configurations.}
\label{fig:light_rings}
\end{figure}

Another source of instability is potentially linked to the presence of light rings. It has been recently shown~\cite{Cunha:2022gde} that exotic ultracompact objects, i.e. compact objects featuring a light ring~\cite{Cardoso:2019rvt}, are unstable under nonlinear perturbations due to the presence of a stable photon orbit in their interior, which traps massless perturbation modes~\cite{Cardoso:2014rha}. Light rings are found as stationary points of the effective potentials 
\begin{equation}
V_{\pm}\propto\frac{-g_{t\varphi}\pm \sqrt{g_{t\varphi}^2-g_{tt}g_{\varphi\varphi}}}{g_{\varphi \varphi}},
\end{equation}
where the plus (minus) sign corresponds to orbits that are co-rotating (counter-rotating) with the star. Among our three configurations, only Case~1 exhibits a pair of light rings (one stable and one unstable) in the effective potential for counter-rotating orbits $V_-$, as depicted in Fig.~\ref{fig:light_rings}. Although this likely makes this configuration prone to instability--~either through migration to a stable non-ultracompact configuration or collapse into a BH~-- the timescale of such instability is unknown due to the absence of simulations for these particular BS models. Furthermore, the frequency of the star is close to the critical frequency for which no light rings are featured, corresponding to an infinite timescale. For these reasons, we chose to keep this solution as an example of high-compactness spinning BS. In any case, stable configurations near Case~1 are expected to have similar geodesic properties. 
\begin{figure*}[htp]
\centering
\includegraphics[width=\textwidth]{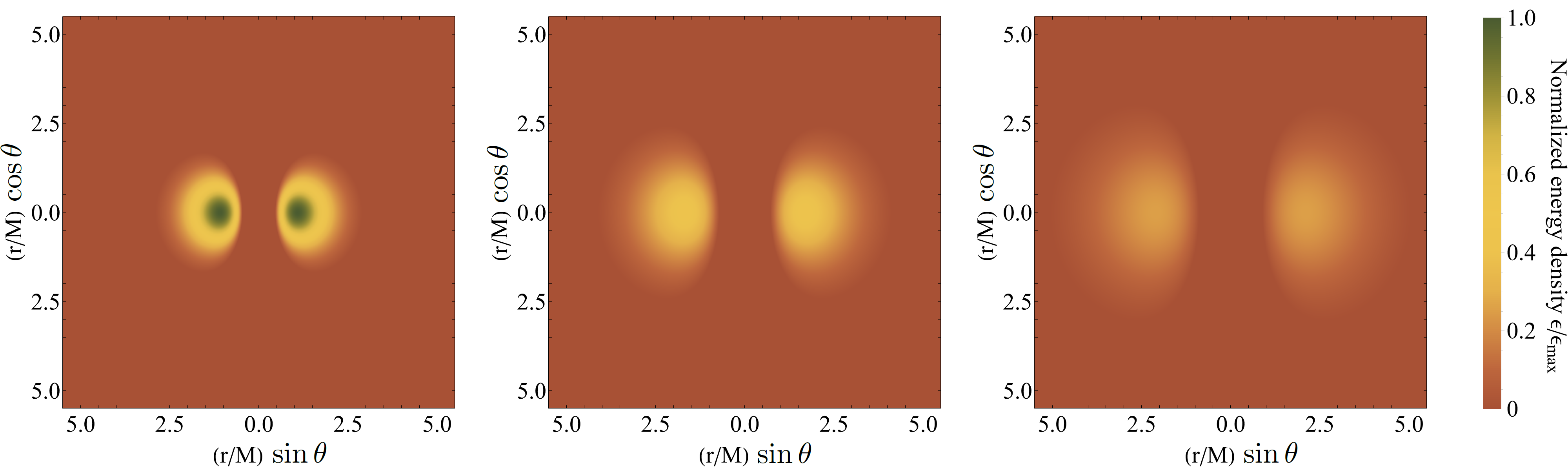}
\caption{Meridional section showing the energy density distribution $\epsilon$ of our representative BSs (from left to right with decreasing compactness: Case~1,~2,~3), normalized to the largest value $\epsilon_\text{max}$ (attained in Case~1).}
\label{fig:density_plots}
\end{figure*}

For completeness, in Fig.~\ref{fig:density_plots}, we present the meridional section of the energy density of the rotating BS cases considered, where the relative compactness and star radii are evident. All energy densities have been normalized with respect to the maximum value reached by Case~1.

\section{Geodesic analysis of generic orbits}\label{sec:geodesics}

Before embarking into the discussion of the results obtained through geodesic evolutions of orbits around rotating BSs, we report that the metric tensor components of our three configurations have been constructed with varying number of grid points on the $r$ and $\theta$ coordinates to test convergence. The islands of stability and corresponding plateau widths at resonance in rotation curves reported below converge to $\lesssim 1\%$ as we increased the number of grid points from $300\times200$ to $1000\times700$. Interestingly, the island and plateau widths converge from below, meaning that more grid points lead to slightly larger islands till convergence occurs, which guarantees their existence regardless of the grid resolution, and therefore ensures nonintegrability. For all simulations presented herein, we find that the constraint equation~\eqref{constraint_equation} is satisfied to within one part in $10^8$ for the first 5 to 10 thousand revolutions (and intersections through the equatorial plane, depending on the initial conditions and BS configuration). 

Finally, in all cases we fix the mass ratio of the EMRI to $m/M=10^{-6}$ and choose the initial energy and $z$-component of the angular momentum as $E/m=0.95$ and $L_z/m=3M$, respectively.

\begin{figure*}[t]
\includegraphics[scale=0.3]{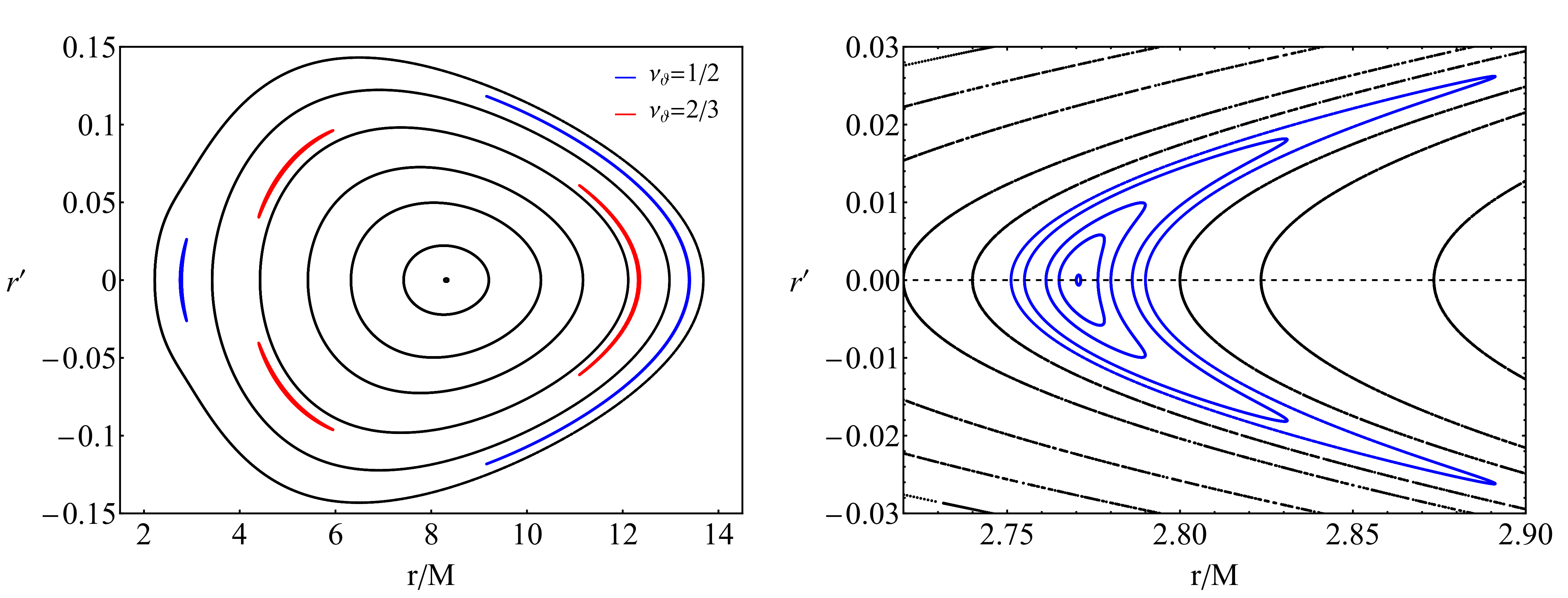}
\caption{Left: Poincar\'e map of a secondary with $m=1M_\odot$ orbiting around a compact rotating BS with $M=10^6M_\odot$, $\mathcal{C}=0.25$ while the rest of the configuration quantities are stated in Table~\ref{tab:configurations} (Case~1). The secondary's conserved energy and angular momentum are $E/m=0.95$, $L_z/m=3M$, respectively. The fixed initial conditions chosen here are $\dot{r}(0)=0,\,\theta(0)=\pi/2$, and $\dot{\theta}(0)$ is defined by the constraint equation to guarantee bound motion, while $r(0)$ is varied. Black curves that surround the central fixed point of the map designate intersections of generic orbits through the equatorial plane, with different initial $r(0)$, while colored curves designate intersections that belong to different resonant islands of stability. Right: Zoom into the leftmost region where the $1/2$ island of stability resides. Similar encapsulated structure is found for the rest of the islands.}\label{fig1}
\end{figure*}
\begin{figure*}[t]
\includegraphics[scale=0.315]{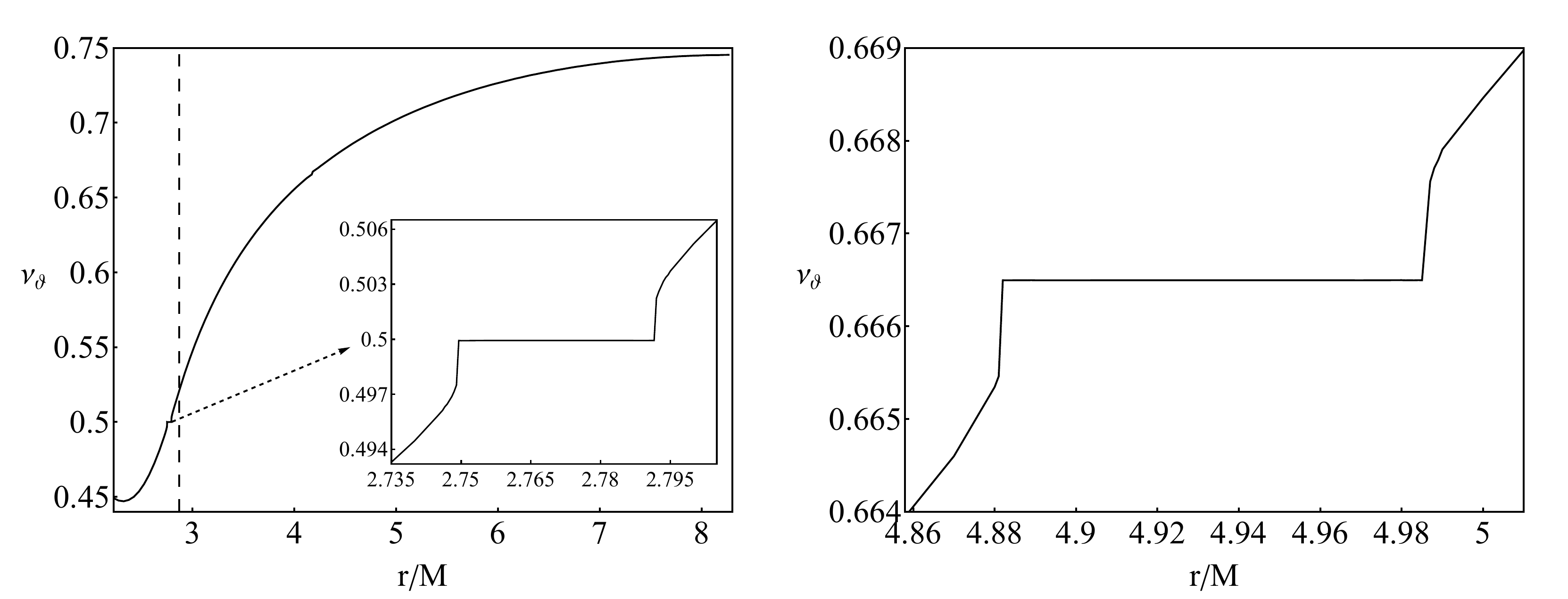}
\caption{Left: Rotation curve corresponding to the same system as in Fig.~\ref{fig1}. The vertical dashed line represents the radius of the BS while the inset zooms in the surrounding region of the $1/2$ plateau. Right: Same as in the left panel but with $\dot{r}(0)=0.1$. The initial velocity gives access to the $2/3$ island of stability.}\label{fig2}
\end{figure*}

\subsection{Case~1}

Case~1 is a representative example for a compact spinning BS. The Poincar\'e map shown in Fig.~\ref{fig1} clearly shows that rotating BSs with large compactness are nonintegrable, since around resonances such as $\nu_\vartheta=1/2,\,2/3$ islands of stability form that encapsulate stable periodic points of geodesics.

From the maps, we calculate the rotation curves for two different choices of initial radial velocity, one with $\dot{r}(0)=r^\prime(0)=0$ and one with $\dot{r}(0)=0.1$, shown in Fig.~\ref{fig2}. The rotation curves for this configuration looks quite similar to those found in~\cite{Lukes-Gerakopoulos:2010ipp,Destounis:2020kss,Destounis:2023gpw}, i.e. an inflection point around the $2/3$ resonance and a plateau at $1/2$ resonance when $\dot{r}(0)=0$. The absence of an event horizon allows for bound geodesics even inside the BS, with its radius designated with a vertical dashed line, surprisingly next to where the $1/2$ plateau resides. Therefore, the first non-BH feature is the existence of resonant islands even inside the star. Another interesting feature is the change in monotonicity close to plunge\footnote{We refer to plunge for the orbits that exit the CZV (the separatrix). In the BH case, these orbits end up plunging into the horizon, while in BSs there might be cases in which plunging orbits escape to infinity.} for Case~1, which has also been observed for Kerr BHs with soft scalar hair and BSs in the study of epicyclic frequencies~\cite{Franchini:2016yvq}.
The inflection point at rotation number $2/3$ for $\dot{r}(0)=0$, in the left panel of Fig.~\ref{fig2}, designates the passage of an orbit between two edges of resonant islands, where there is an unstable periodic point. Changing the initial radial velocity to $\dot{r}(0)=0.1$, as we show for the rotation curve in Fig.~\ref{fig2} (right panel), the phase space trajectory crosses the island and a plateau at $2/3$ appears.

\begin{figure*}[t]
\includegraphics[scale=0.3]{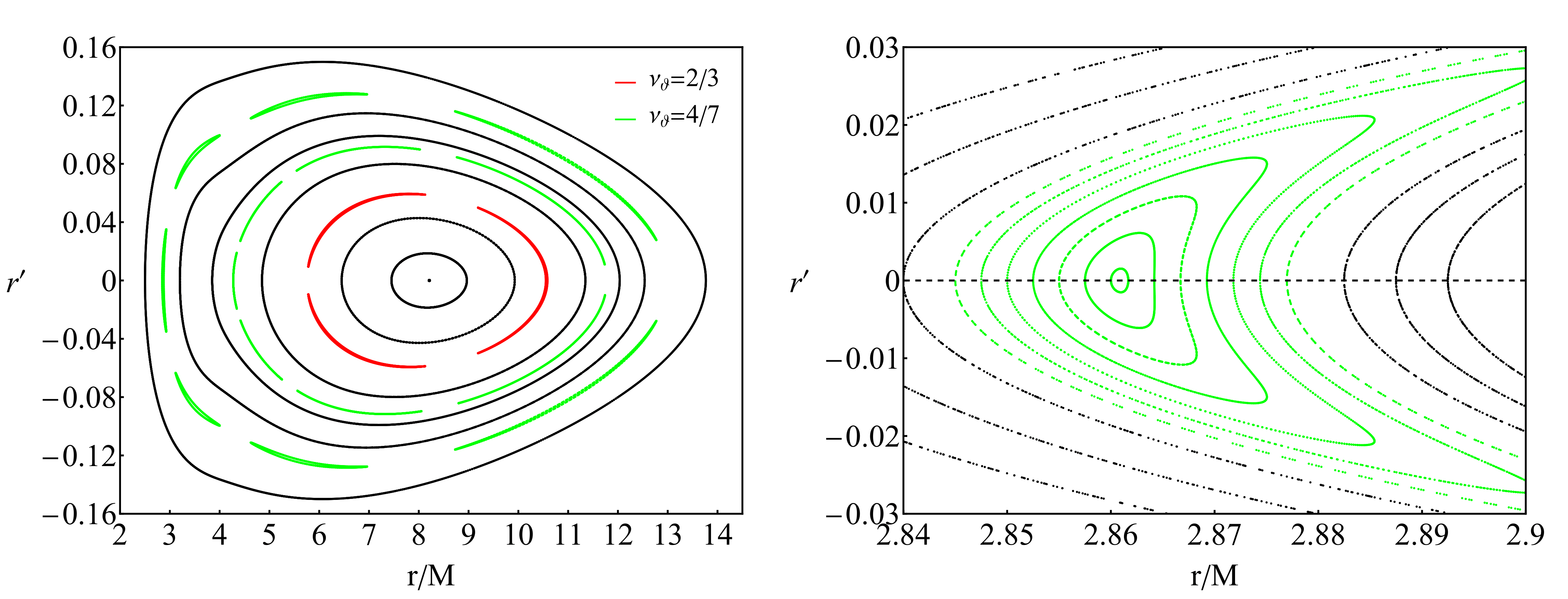}
\caption{Left: Same as in Fig.~\ref{fig1} but for Case~2 (intermediate compactness $\mathcal{C}=0.19$). Right: Zoom into the leftmost region where the $4/7$ island of stability resides. Similar encapsulated structure is found for the rest of the islands.}\label{fig3}
\end{figure*}
\begin{figure*}[t]
\includegraphics[scale=0.3]{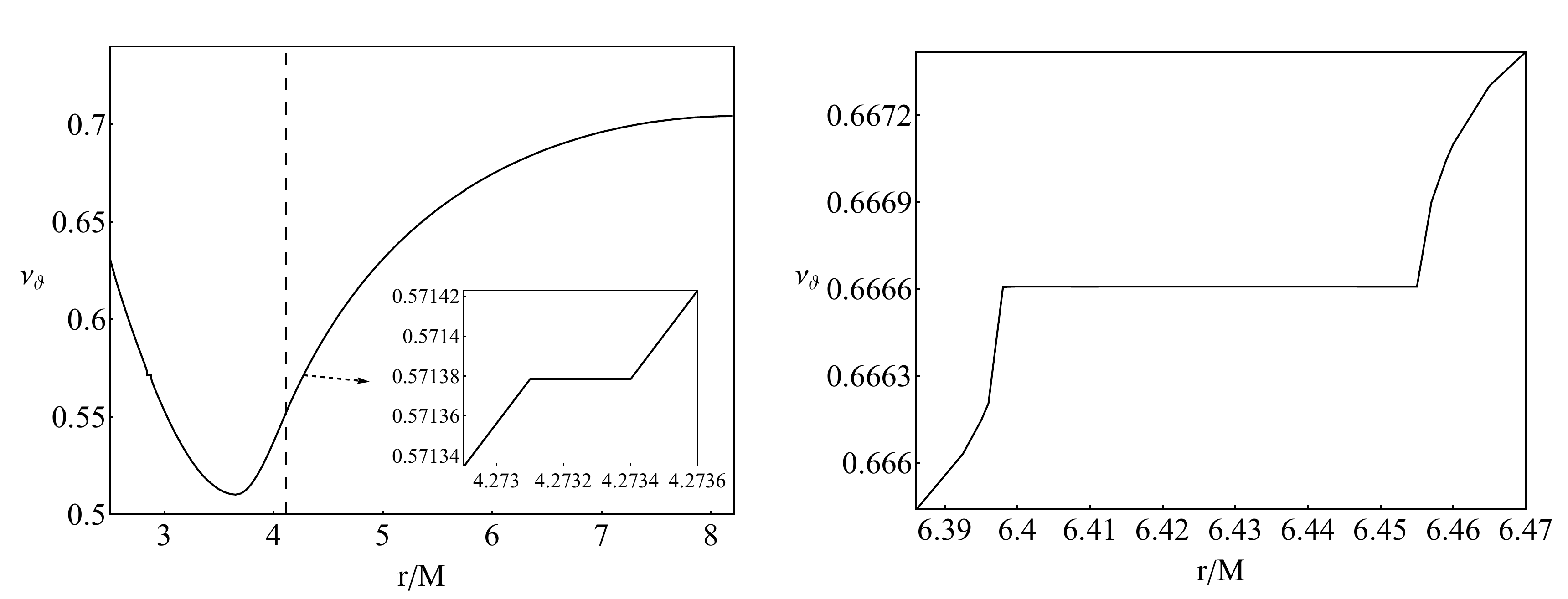}
\caption{Left: Rotation curve corresponding to the same system as in Fig.~\ref{fig2}. The horizontal dashed line represents the radius of the BS while the inset zooms in the surrounding region of the external $4/7$ plateau. Right: Same as in the left panel but with $\dot{r}(0)=0.058$. The initial velocity gives access to the $2/3$ island of stability.}\label{fig4}
\end{figure*}

\subsection{Case~2}

As we slightly decrease the compactness of the BS, the non-BH effects are further enhanced. For an intermediate compactness rotating BS (Case~2) the Poincar\'e map, shown in Fig.~\ref{fig3}, displays a much more intricate structure around subdominant resonances. Firstly, the radius of the star increases, therefore orbits are allowed to exist in its interior, which is precluded for BHs, and should lead to more prominent effects. Intriguingly, due to the aforementioned phenomenon, here we find for the first time degenerate resonant islands --~namely islands of stability occurring for two different ranges of the radial coordinate. In particular, Fig.~\ref{fig3} shows the typical exterior $2/3$ island as well as two sets of Birkhoff chains for the $4/7$ resonance, one in the exterior and one in the interior of the BS.

The rotation curve in Fig.~\ref{fig4} for $\dot{r}(0)=0$ (left panel) confirms all the above. Decreasing the compactness leads to a more prominent change in monotonicity when the secondary enters the star. This in turn allows for degenerate plateaus both in the interior and the exterior of the star's geodesics, with the interior plateau being wider than the exterior one. Again, choosing a different initial velocity, we can turn the inflection point of $2/3$ resonance into a plateau, since the velocity pushes the orbit to traverse the island.

\begin{figure*}[t]
\includegraphics[scale=0.3]{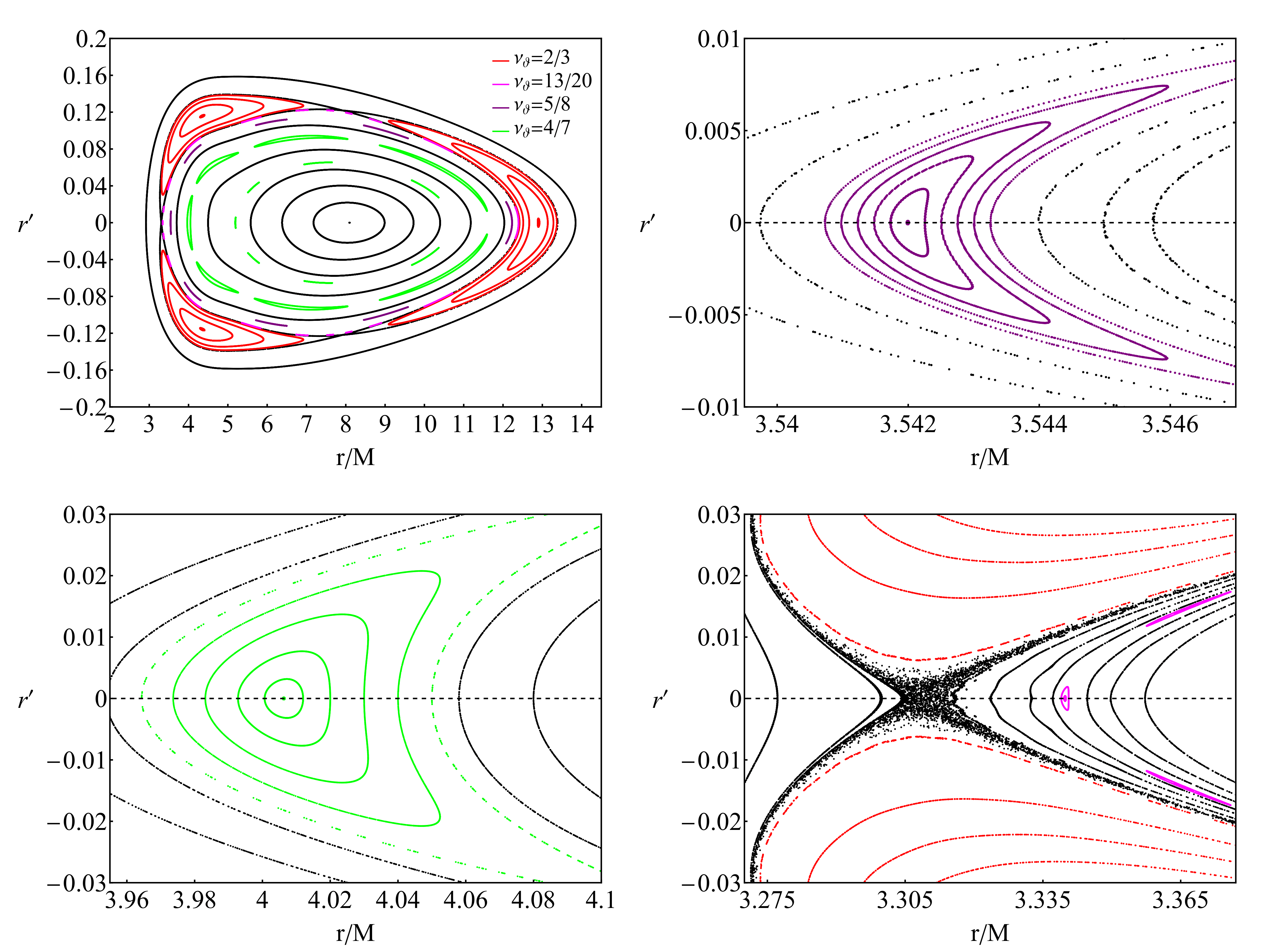}
\caption{Top left:  Same as in Figs.~\ref{fig1} and~\ref{fig2} but for Case~3 (smaller compactness, $\mathcal{C}=0.16$). Top right: Zoom into the region where the leftmost $5/8$ island of stability resides. Bottom left: Zoom into the region where the leftmost $4/7$ island of stability resides. Bottom right: Zoom into a region of the Poincar\'e map on the top left where a chaotic layer is present, shown with scattered black points that emanate from the unstable $2/3$ periodic point with $\dot{r}(0)=0$, and parts of two islands of stability, namely the $2/3$ and $13/20$.}\label{fig5}
\end{figure*}
\begin{figure*}[t]
\includegraphics[scale=0.31]{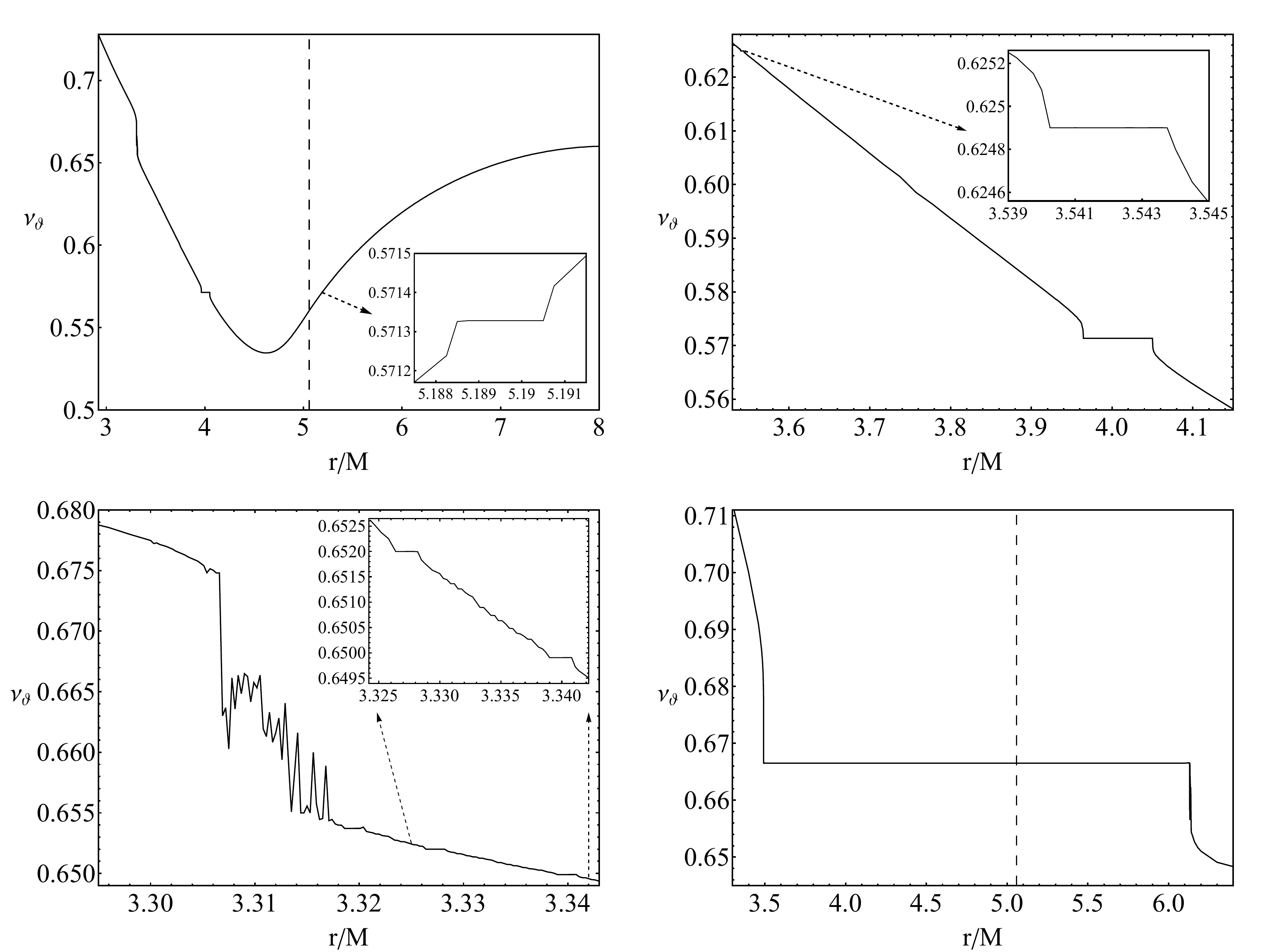}
\caption{Top left: Rotation curve corresponding to the same system as in Fig.~\ref{fig5}. The vertical dashed line represents the radius of the BS while the inset zooms in the surrounding region of the external $4/7$ plateau. Top right: Zoomed part of the rotation curve on the top left figure on the subdominant resonances $\nu_\vartheta=5/8$ and $4/7$ in descending order. Bottom left: Zoomed part of the rotation curve on top left figure at the chaotic layer of the unstable periodic point for $2/3$ resonance. A further zoom is shown in the inset of two extremely subdominant islands with $\nu_\vartheta=15/23$ and $13/20$ in descending order. Bottom right:  Rotation curve for the same parameters and initial conditions as in the top left figure but with $\dot{r}(0)=0.155$. The initial velocity gives access to the $2/3$ island of stability.}\label{fig6}
\end{figure*}
 
\subsection{Case~3}

Case~3 is probably the most interesting one. Its relatively smaller compactness allows for a plethora of new phenomena in both Poincar\'e maps (Fig.~\ref{fig5}) and rotation curves (Fig.~\ref{fig6}). The radius of the BS increases even more and the CZV enters deep inside the star where bound geodesics are still possible. This leads to a multitude of dominant and subdominant resonances, as shown in the top left panel of Fig.~\ref{fig5}, as well as further degenerate Birkhoff chains. In the top right and bottom left panels, a zoom of the interior $5/8$ and $4/7$ islands is presented. Probably the most interesting feature is displayed in the bottom right panel. Two islands of stability appear, namely the $2/3$ (in red) and the interior $13/20$ part of the island (in pink). Furthermore, we find a visible chaotic layer that surrounds the $2/3$ island, designated with scattered black points. Their source is the unstable periodic point at $\dot{r}(0)=0$ and $r(0)\sim3.31 M$. To the best of our knowledge, this is the first case in which a chaotic layer appears for a motivated model of compact object without any pathologies. This layer should give rise to significant effects on rotation curves and eventual EMRIs crossing through the particular $2/3$ island when the fluxes will be taken into account~\cite{Destounis:2021mqv,Destounis:2021rko,Destounis:2023gpw}.

Figure~\ref{fig6} shows the significantly modified rotation curves with respect to a typical non-Kerr BH (see, e.g.,~\cite{Apostolatos:2009vu,Lukes-Gerakopoulos:2010ipp,Contopoulos:2011dz,Lukes-Gerakopoulos:2012qpc,Zelenka:2017aqn,Destounis:2020kss,Lukes-Gerakopoulos:2021ybx,Deich:2022vna,Chen:2022znf,Destounis:2023gpw}). Degenerate $4/7$ plateaus are visible in the interior and exterior of the star, as well as a robust internal plateau for the $5/8$ resonance. Although finding its external counterpart requires extreme fine-tuning, it is guaranteed that such region exists in the exterior. Closing in to the plunge (i.e., the inner boundary of the CZV), deep inside the star, a very large inflection point appears, which after zooming in (see Fig.~\ref{fig6}, bottom left) reveals a thick chaotic layer with ill-defined rotation numbers. We also find a couple of extremely subdominant islands for the $13/20$ (also shown in Fig.~\ref{fig5}) and $15/23$ resonances. Even more interestingly, by assuming an appropriate initial velocity for the geodesics, we managed to find, for the first time, an island (which leads to a plateau) that begins from the exterior of the star and ends in the interior. Its width ($\sim2.6 M$ in quasi-isotropic or $\sim 3.6M$ in Boyer-Lindquist coordinates) is so wide-spread that supersedes any other plateau ever found in non-Kerr spacetimes, where the widest one found is of order $\sim0.05M$ (in Boyer-Lindquist coordinates) and barely compares with the plateaus for Case~1 ($2/3$ resonance) and Case~2 ($4/7$ resonance) island widths. Nevertheless, we need to point out that the width of each island found in this and other similar studies are initial-condition dependent, according to their multiplicity. Therefore, maximizing their width is a tedious task. A better method would be to fix an initial condition, such as $\dot{r}(0)=0$, and cross the islands forming beyond the central periodic point of the Poincar\'e map. In a geodesic analysis this is perfectly doable in normal timescales and without accumulating significant error during the orbital evolution, though an inspiral trajectory close to the plunge is much faster to produce with minimized numerical error, in contrast to the islands on the right side of the central point where the fluxes are much smaller and the evolution time needs to be increased a lot, which in turn gives rise to larger errors.

\begin{figure}[t]
\includegraphics[scale=0.315]{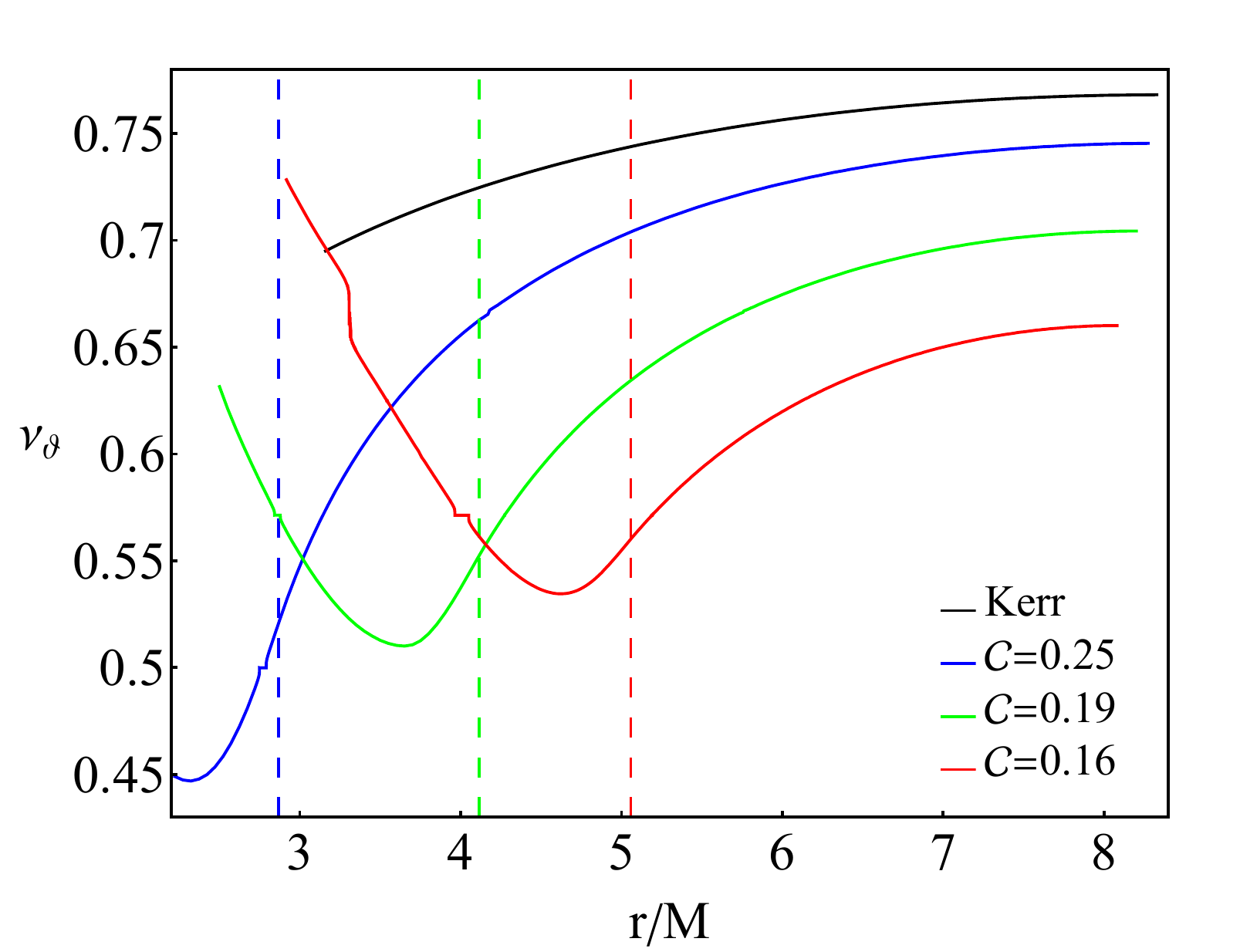}
\caption{Combined rotation curves of a secondary with $m=1M_\odot$ orbiting around a rotating BS with $M=10^6M_\odot$, varying compactness $\mathcal{C}$ and $\chi=0.8$. The secondary's conserved energy and angular momentum are $E/m=0.95$, $L_z/m=3M$, respectively. The fixed initial conditions chosen here are $\dot{r}(0)=0,\,\theta(0)=\pi/2$ and $\dot{\theta}(0)$ is defined by the constraint equation to guarantee bound motion, while $r(0)$ is varied. The horizontal dashed lines represents the radii of each BS configuration. For comparison, we include the rotation curve of a Kerr BH with the same spacetime and geodesic parameters.}\label{fig7}
\end{figure}

For completeness, in Fig.~\ref{fig7} we present all rotation curves obtained with $\dot{r}(0)=0$ and varying $r/M$, together with the rotation curve of a Kerr BH with $\chi=0.8$. The differences between BH orbits and non-compact rotating BSs are evident and the BSs present novel features due to the absence of an event horizon. For fixed $E$ and $L_z$, we find that even the most compact configuration considered differs dramatically from that of Kerr, while the rest completely disengage with the typical behavior of a Kerr rotation curve. Perhaps a rotation curve in the lines of those presented in~\cite{Lukes-Gerakopoulos:2010ipp,Destounis:2020kss} may seem more similar to that of Case~1 BS but still the existence of the radius, the change in monotonicity when the orbits enters the star, the degenerate plateaus, and the existence of an observable chaotic layer should make rotating BSs having a contrasting and discernible behavior of geodesics when compared to those of non-Kerr spacetimes.

\section{Adiabatic inspiral and waveforms}\label{sec:inspiral}
Inspiral in the EMRI limit can be conveniently described within BH perturbation theory~\cite{Barack:2009ux,Pound:2021qin}. Owing to the hierarchy of scales, the dynamics can be studied as a small perturbation of the geodesic motion of the secondary test mass around the primary object. To leading order in the mass ratio, one could evolve geodesic quantities adiabatically by taking into account dissipation due to radiative degrees of freedom. Higher order corrections require including conservative and, in general, self-force effects during the inspiral~\cite{Barack:2009ux,vandeMeent:2016pee,vandeMeent:2017bcc,Pound:2017psq,Pound:2021qin}.

While this program has been extremely successful for standard EMRIs around a Kerr BH within General Relativity, going beyond the standard paradigm is much more challenging, even at the leading order. Indeed, dissipative corrections are computed using the Teukolsky formalism~\cite{Teukolsky:1973,Teukolsky:1974yv,Detweiler:1978ge}, which allows separating the perturbations of a Kerr BH in General Relativity and computing fluxes through numerically integration of inhomogeneous ordinary differential equations with a point-particle source term~\cite{Sasaki:1981sx,Mino:1997bx}. This technology has been widely tested in the frequency~\cite{Hughes:1999bq,Hughes:2001jr,Drasco:2005kz} and time domain~\cite{Poisson:2004cw,Martel:2003jj,Lopez-Aleman:2003sib,Khanna:2003qv}. However, it heavily relies on the separability of the perturbation equations, which does not occur if the background is not described by the Kerr metric as in our case (which is, in addition, only known numerically). 
Likewise, the gravitational self-force was computed up to second-order in perturbations~\cite{Pound:2012nt,Pound:2019lzj,Pound:2021qin} for orbits around BHs~\cite{vandeMeent:2017bcc,2620937}, but the case of ECOs is an uncharted territory.

In the absence of a consistent framework to study EMRI dynamics around a non-Kerr spinning object, here we use the only framework available at the moment, namely approximate semi-relativistic inspirals~\cite{Glampedakis:2002cb,Gair:2005ih} and waveforms with methods known as ``kludge'' schemes~\cite{Ruffini:1981,Sasaki:1983,Babak:2006uv}. Kludge waveforms are constructed through the so-called numerical kludge scheme, namely by combining flat spacetime weak-field (PN) GW emission together with a fully-relativistic treatment for the secondary's motion.
	
\subsection{Numerical kludge scheme}
	
To approximate EMRIs around rotating BSs we integrate the second-order dynamical system for $r,\,\theta$, augmented with weak-field PN fluxes for the energy and angular momentum loss due to GW emission~\cite{Glampedakis:2002ya,Glampedakis:2002cb,Gair:2005ih}. This treatment, though approximate and valid only for small orbital velocities, takes into account the dominant contribution of the secondary's radiative backreaction to the primary's geometry, at second PN order, and results in an adiabatic evolution of the EMRI. During the evolution, the orbit is treated, at small timescales, as a geodesic, while for longer timescales the trajectory is slowly driven through consecutively damped geodesics. This scheme has been shown to perform well when compared to Teukolsky-based waveforms of EMRIs~\cite{Babak:2006uv}. 
	
BSs have non-trivial multipolar structure which differs from the one of a Kerr BH~\cite{Ryan:1995wh,Ryan:1997hg,Ryan:1997kh,Pacilio:2020jza,Vaglio:2022flq,Vaglio:2023lrd}. At second PN order, the kludge scheme~\cite{Glampedakis:2002cb} involves the mass quadrupole moment $M_2$. Thus, to construct a more faithful (though still approximate) inspiral around a rotating BS, we augment the fluxes with its modified mass quadrupole moment (see~\cite{Barack:2006pq,Gair:2007kr,Apostolatos:2009vu,Lukes-Gerakopoulos:2010ipp,Destounis:2021mqv,Destounis:2021rko}). This kludge scheme, together with the modified mass quadrupole moment, has recently been examined and found to provide results qualitatively equivalent to evolutions with instantaneous self-force in non-Kerr electromagnetic analogs, which indicates that this method can, in principle, describe resonance and island-crossings in nonintegrable EMRIs with sufficient accuracy~\cite{Mukherjee:2022dju}.
	
We employ linear variations of $E$ and $L_z$ in an iterative way, such that~\cite{Canizares:2012is,Destounis:2021mqv,Destounis:2021rko}
\begin{align}
\label{update_E}
E_1&=\frac{E(0)}{m}+\bigg< \frac{dE}{dt}\bigg>\bigg|_{t=0} N_r\, T_r,\\
\label{update_L}
L_{z,1}&=\frac{L_{z}(0)}{m}+ \bigg< \frac{dL_z}{dt}\bigg>\bigg|_{t=0} N_r\, T_r,
\end{align}
where $E(0),\,L_{z}(0)$ are the energy and $z$-component of the angular momentum at $t=0$, respectively. In turn, $\langle{dE}/{dt}\rangle|_{t=0},\,\langle{dL_z}/{dt}\rangle|_{t=0}$ are the averaged PN fluxes calculated at the beginning of the inspiral, through the complicated equations outlined in~\cite{Gair:2005ih,Destounis:2021rko}. $N_r$ is the number of radial periods elapsed between each update of~\eqref{update_E} and~\eqref{update_L}, which improves the scheme and includes cumulative nonlinear variations, while $T_r$ is the radial period of the EMRI. Equations~\eqref{update_E} and~\eqref{update_L} are iterated along the whole EMRI evolution with appropriate choices of $N_r$ and $T_r$ to obtain the dissipative orbit. If the system was integrable, then through the evolution of the Carter constant, we would be able to also evolve the rest of the components of the angular momentum, namely $L_x$ and $L_y$, which are  imprinted in Carter's constant. Our case though is nonintegrable and lacks of a separation Carter-like constant; therefore, we have no explicit way of evolving $L_x$ and $L_y$ but rather we keep them constant.
	
Even though, due to dissipation, the constraint equation~\eqref{constraint_equation} is not anymore a constant of motion (that can be monitored to assess the accuracy of the numerical scheme), we have tested numerous random time instants of the inspirals obtained with the aforementioned scheme from which we extract the instantaneous position and velocity vectors of the orbit, $E$ and $L_z$, and calculate~\eqref{constraint_equation} for these values. We then compare the resulting constraint with the one obtained by evolving a geodesic with the aforementioned parameters of the same time instants as initial conditions. For all cases, we find that the new dissipated constraint is satisfied to within 1 part in $10^7$ for the first $\sim10^4$ revolutions. Through geodesic evolutions of successive time instants of an inspiral, we can built dissipative Poincare maps with adiabatically decreasing $E$ and $L_z$ and therefore calculate time-dependent rotation curves~\cite{Destounis:2021rko,Lukes-Gerakopoulos:2021ybx}.

\begin{figure*}[t]
\includegraphics[scale=0.3]{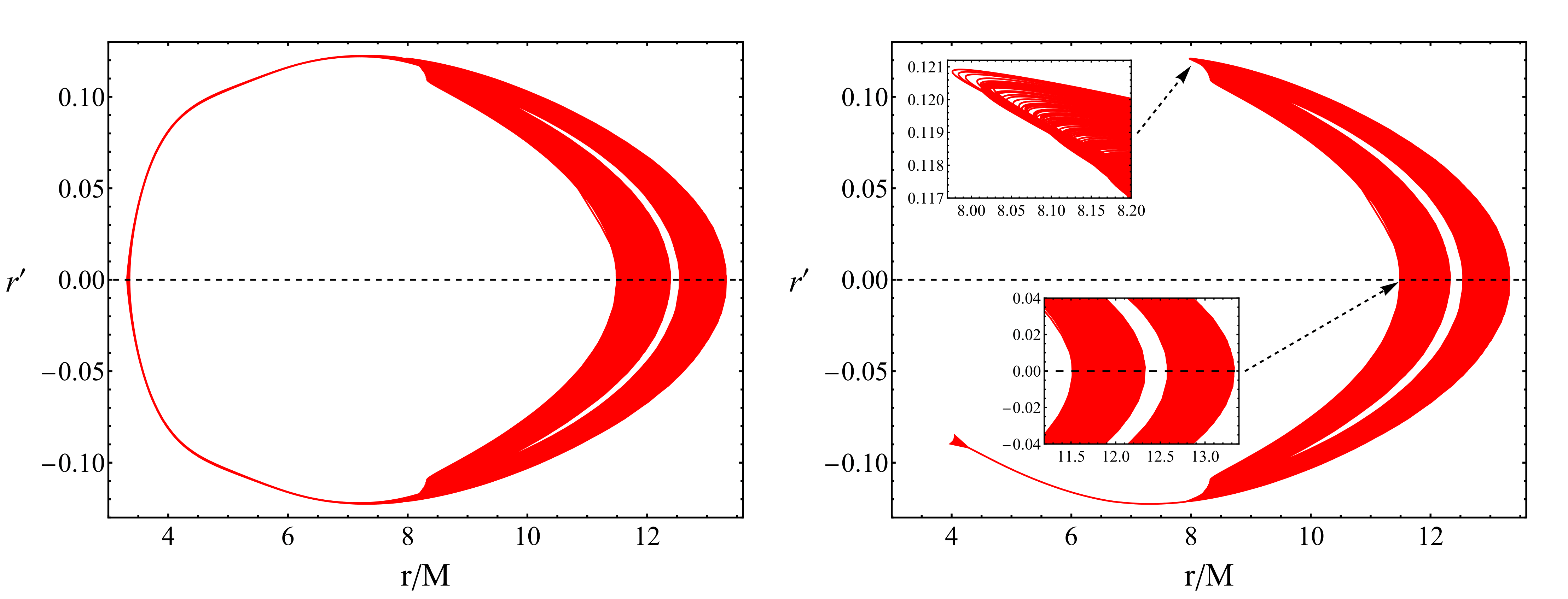}
\caption{Left: Sustained resonance for Case~3, with mass ratio $m/M=10^{-6}$, and simplified linear fluxes without any updates as given in Eqs.~\eqref{update_E} and~\eqref{update_L}. The secondary's initial parameters and conditions are $E(0)/m=0.95$, $L_z(0)/m=3M$, $r(0)=6.5M$, $\dot{r}(0)=0.155$, $\theta(0)=\pi/2$, while $\dot{\theta}(0)$ is defined from the constraint Eq.~\eqref{constraint_equation}. The total evolution time is $t=5\times10^6 M\sim 10$ months. The linearly dissipative inspiral spends roughly $1$ month off resonance and $9$ months in perfect resonance. Here, we have chosen a stroboscopic depiction for the trapping to be easier identified, i.e. from the time series of the
full Poincar\'e map, only every third consequent point is kept. Right: Same as in left panel but starting from a point of the inspiral where the $1$ month off resonance have lapsed and the orbit enters the (putatively) eternal resonance.}\label{fig8}
\end{figure*}

\subsection{GW modeling}
	
To present the phenomenological imprints of inspiraling secondaries onto supermassive BSs, we employ the quadrupole approximation. In the traceless and tranverse gauge the metric perturbations read (e.g.,~\cite{Misner:1973prb})
\begin{equation}\label{metpert}
h_{ij}=\frac{2}{d}\frac{d^2 Q_{ij}}{dt^2},
\end{equation}
where $Q_{ij}$ is the symmetric and trace-free (STF) mass quadrupole tensor, which can be written as
\begin{equation}
Q^{ij}=\left[\int x^i x^j T^{tt}(t,x^k) \,d^3 x\right]^\text{STF},
\end{equation}
with $t$ the coordinate time of the secondary source, and
\begin{equation}\label{Ttt}
T^{tt}(t,x^i)=m \delta^{(3)}\left[x^i-Z^i(t)\right].
\end{equation}
Here, we employ the  approximation where quasi-isotropic coordinates at infinity are identified with spherical ones as $Z(t)=(x(t),y(t),z(t))$, where
\begin{align}
x(t)&=r(t)\sin\theta(t)\cos\phi(t),\\
y(t)&=r(t)\sin\theta(t)\sin\phi(t),\\
z(t)&=r(t)\cos\theta(t),
\end{align} 
and then transform them from spherical to Euclidean coordinates that describe the secondary's trajectory.
	
An incoming wave from an EMRI onto an interferometer can be projected in two polarizations, $+$ and $\times$, by introducing two unit vectors $\boldsymbol{\hat{p}}=\boldsymbol{\hat{n}}\times\boldsymbol{\hat{z}}/|\boldsymbol{\hat{n}}\times\boldsymbol{\hat{z}}|$ and $\boldsymbol{\hat{q}}=\boldsymbol{\hat{p}}\times\boldsymbol{\hat{n}}$ (here $\times$ is the cross product of two vectors and should not be confused with the cross polarization symbol of the incoming GW), which are defined in terms of a third unit vector $\boldsymbol{\hat{n}}$ that points from the EMRI source to the detector. Finally, the unit vector $\boldsymbol{\hat{z}}$ designates the spin direction of the BS. The triplet $\boldsymbol{\hat{p}},\,\boldsymbol{\hat{q}},\,\boldsymbol{\hat{n}}$ forms an orthonormal basis from which the polarization tensor components are defined as
\begin{equation}
\epsilon_+^{ij}=p^i p^j-q^i q^j,\,\,\,\,\,\,\epsilon_\times^{ij}=p^i q^j+p^j q^i,
\end{equation}
and allow us to write the metric perturbation in the quadrupole approximation as
\begin{equation}
h^{ij}(t)=\epsilon_+^{ij}h_+(t)+\epsilon_\times^{ij}h_\times(t),
\end{equation}
where $h_{+,\times}$ are the plus and cross polarizations of the incoming GW. The GW components are then expressed in terms of the position, $Z^i(t)$, velocity, $v^i(t)=dZ^i/dt$, and acceleration, $a^i(t)=d^2Z^i/dt^2$, vectors. One finally obtains
\begin{equation}
\label{GW_formula}
h_{+,\times}(t)=\frac{2 m}{d}\epsilon^{+,\times}_{ij}\left[a^i(t)Z^j(t)+v^i(t)v^j(t)\right].
\end{equation}
	
LISA's response to an incident GW event is correlated with the antenna pattern functions, $F^{+,\times}_{I,II}(t)$, describing the motion of the detector on its respective spacecraft channels $I$ and $II$ (see Refs.~\cite{Cutler:1997ta,Barack:2003fp,Destounis:2020kss} for details). The total waveform detected by a LISA-like interferometer reads
\begin{equation}\label{total_GW}
h_{I,II}(t)=\frac{\sqrt{3}}{2}\left[F^+_{I,II}(t) h_+(t)+F^\times_{I,II}(t) h_\times(t)\right],
\end{equation}
We assume a detector that lies at a luminosity distance $d$ with fixed orientation $\boldsymbol{n}=(0,0,1)$ with respect to the EMRI source, and that the primary's polar and azimuthal angles are fixed at the equatorial plane. The data streams that will be considered in what follows will contain the GW together with stationary and Gaussian noise. For simplicity, we will further assume that the two data stream channels are uncorrelated, thus we will abide to a single-channel approximation of our detector.

\section{EMRIs around a spinning BS}

In this section, we discuss the effects of non-integrability and chaos on EMRI evolution and GW emission. Even though approximate, the results shown below demonstrate the basic features of the chaotic phenomena taking place in rotating BSs. Since the most interesting cases of the BSs constructed are Cases~1 and 3, we will focus on them from now on. We fix $E(0)/m=0.95$, $L_z(0)/m=3M$ as initial parameters of the secondary in Eqs.~\eqref{update_E} and~\eqref{update_L}, respectively, and perform a different numbers of updates depending on how quick the inspiral evolves, namely how close we are to the inner boundary of the CZV (separatrix).

Note that in some cases the inspiral would occur \emph{inside} the BS. This is a striking difference with respect to the BH case in which the signal disappears after the particle has crossed the horizon. We can, therefore, continue the evolution as long as the orbits do not cross the separatrix, where the inspiral plunges.

It is worth noticing that the potential inside the BS is approximately constant (see Fig.~\ref{fig:potential}) and not particularly strong, which justifies the use of PN fluxes even inside the star. For EMRIs around Kerr, the overlap between PN and Teukolsky-based waveforms starts to completely deteriorate typically when the periapsis radius $r_p\sim 5M$ (in Boyer-Lindquist coordinates) or smaller~\cite{Babak:2006uv}, but in the BH case the potential is stronger, as shown in Fig.~\ref{fig:potential}, due to the larger compactness at the horizon. Thus, our PN approximation should be accurate enough also inside the star all the way to the edge of the CZV, especially for the least compact configurations. Furthermore, we stress that we are only considering radiation-reaction effects, neglecting environmental effects such as dynamical friction and accretion within the BS~\cite{Macedo:2013jja}, as well as direct (non-gravitational) coupling between the secondary and the scalar field, both of which might significantly contribute to the inspiral.

\subsection{Sustained resonances}

A first interesting result arises when we ignore the updates on the flux, Eqs.~\eqref{update_E} and~\eqref{update_L}, an approximation that has been considered in various analyses~\cite{Apostolatos:2009vu,Lukes-Gerakopoulos:2010ipp,Lukes-Gerakopoulos:2012qpc}. This simplification leads to the following linearly-variated fluxes 
\begin{align}
    \label{E_linear}
E(t)&=\frac{E(0)}{m}+\bigg< \frac{dE}{dt}\bigg>\bigg|_{t=0} t,\\
\label{Lz_linear}
L_{z}(t)&=\frac{L_{z}(0)}{m}+ \bigg< \frac{dL_z}{dt}\bigg>\bigg|_{t=0} t,
\end{align}
Such assumption inevitably leads to sustained resonances, that cannot exist in Kerr~\cite{vandeMeent:2013sza}, where the secondary is trapped in a resonant island for an extremely long (potentially infinite) time. In Fig.~\ref{fig8} we present one such case, where a $10$ month inspiral spends $1$ month evolving normally and $9$ months in perfect resonance. It never eventually escapes the island, at least for the timescale over we have evolved the EMRI. In order to identify the trapping more easily, Fig.~\ref{fig8} presents a stroboscopic depiction, i.e. from the time series of the full dissipative Poincar\'e map, only every third consequent point is kept due to the multiplicity of the $2/3$ resonance.

The initial position for a sustained resonance to occur is not fine-tuned but rather a quite large region of initial conditions $r(0)\in[6.46,6.58]M$ exists (which corresponds to $[7.47,7.59]M$ in Boyer-Lindquist coordinates), when fixing the rest of the parameters as $\dot{r}(0)=0.155$, $\theta(0)=\pi/2$, and define $\dot{\theta}(0)$ through the constraint equation~\eqref{constraint_equation}. Similar results have recently been found in~\cite{Lukes-Gerakopoulos:2021ybx} for linear variations of the fluxes like Eq.~\eqref{E_linear}.
Even if the range of initial conditions giving such sustained resonances is not negligible, in practice such sustained resonances are an artifact due the assumption of low-order energy and angular-momentum fluxes throughout the evolution. Namely, the linear approximation in energy and angular momentum is valid as long as the trajectory of the EMRI does not get too far from the initial geodesic, on which the orbital elements and fluxes have been initialized. 

\begin{figure}[t]
\includegraphics[scale=0.31]{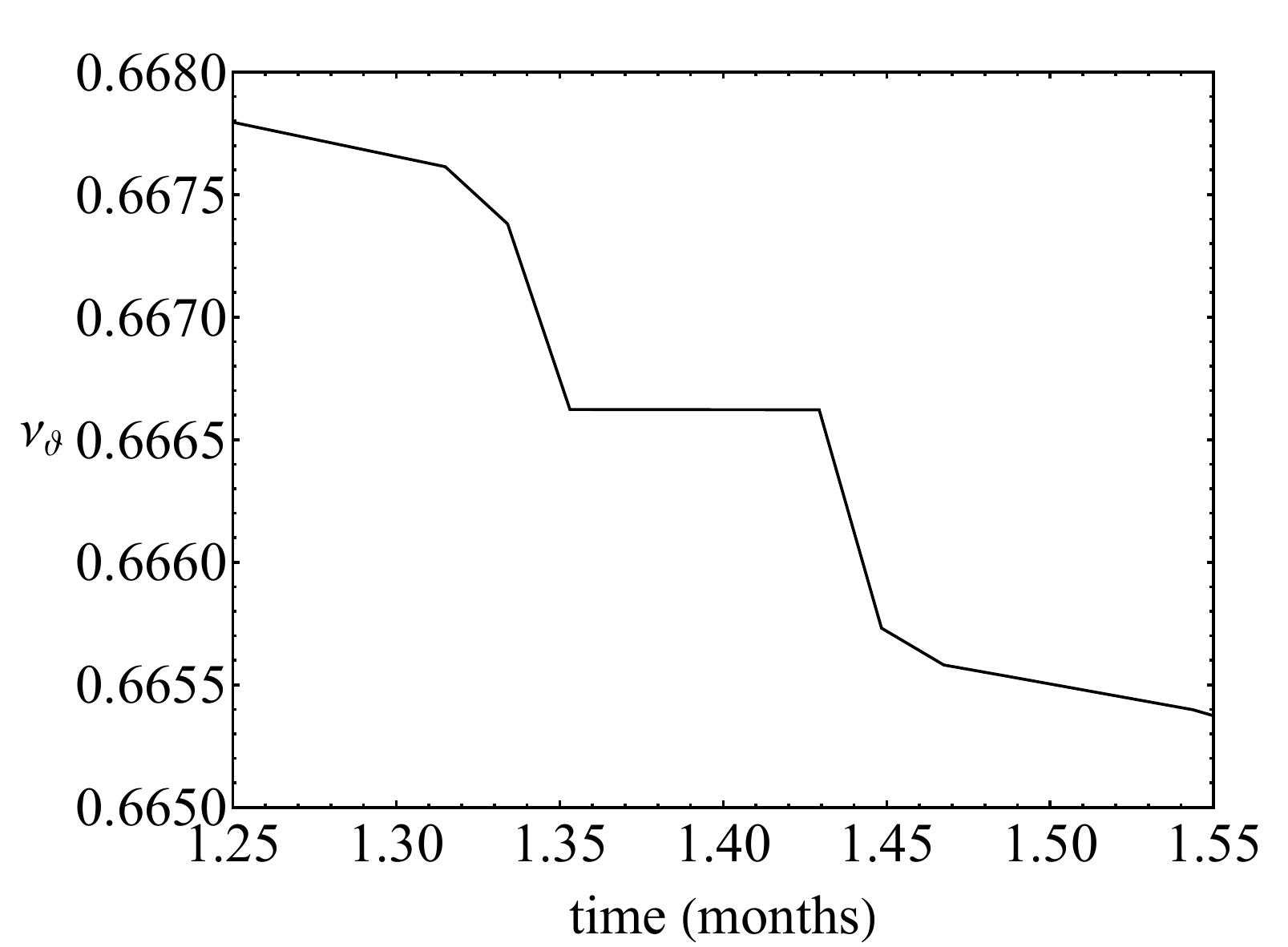}
\caption{External $2/3$ resonant island crossing for Case~1. To produce the inspiral we have updated the fluxes $300$ times. The secondary's initial parameters and conditions are $E(0)/m=0.95$, $L_z(0)/m=3M$, $r(0)=5.026M$, $\dot{r}(0)=0.1$, $\theta(0)=\pi/2$, while $\dot{\theta}(0)$ is defined from the constraint Eq.~\eqref{constraint_equation}. The EMRI spends $\sim200$ cycles in resonance.}\label{fig9}
\end{figure}

By taking into account updates (from $50$ to $150$ for some representative cases) on the fluxes for the aforementioned radial range, which effectively includes nonlinear terms on the fluxes, all sustained resonances vanish. Of course, we cannot exclude the possibility that the region of sustained resonances either shrinks significantly or hides into a different range of initial conditions, although we consider this option as unlikely. Note also that putative sustained resonances should persist (and not disappearing) as the number of flux updates is increased. A similar case was recently studied for an EMRI analog  in~\cite{Mukherjee:2022dju}, where lower order flux approximations gave rise to sustained resonances whereas, when higher order terms were involved, the sustained resonances disappeared.

\subsection{Time-dependent rotation curves}

\begin{figure}[t]
\includegraphics[scale=0.31]{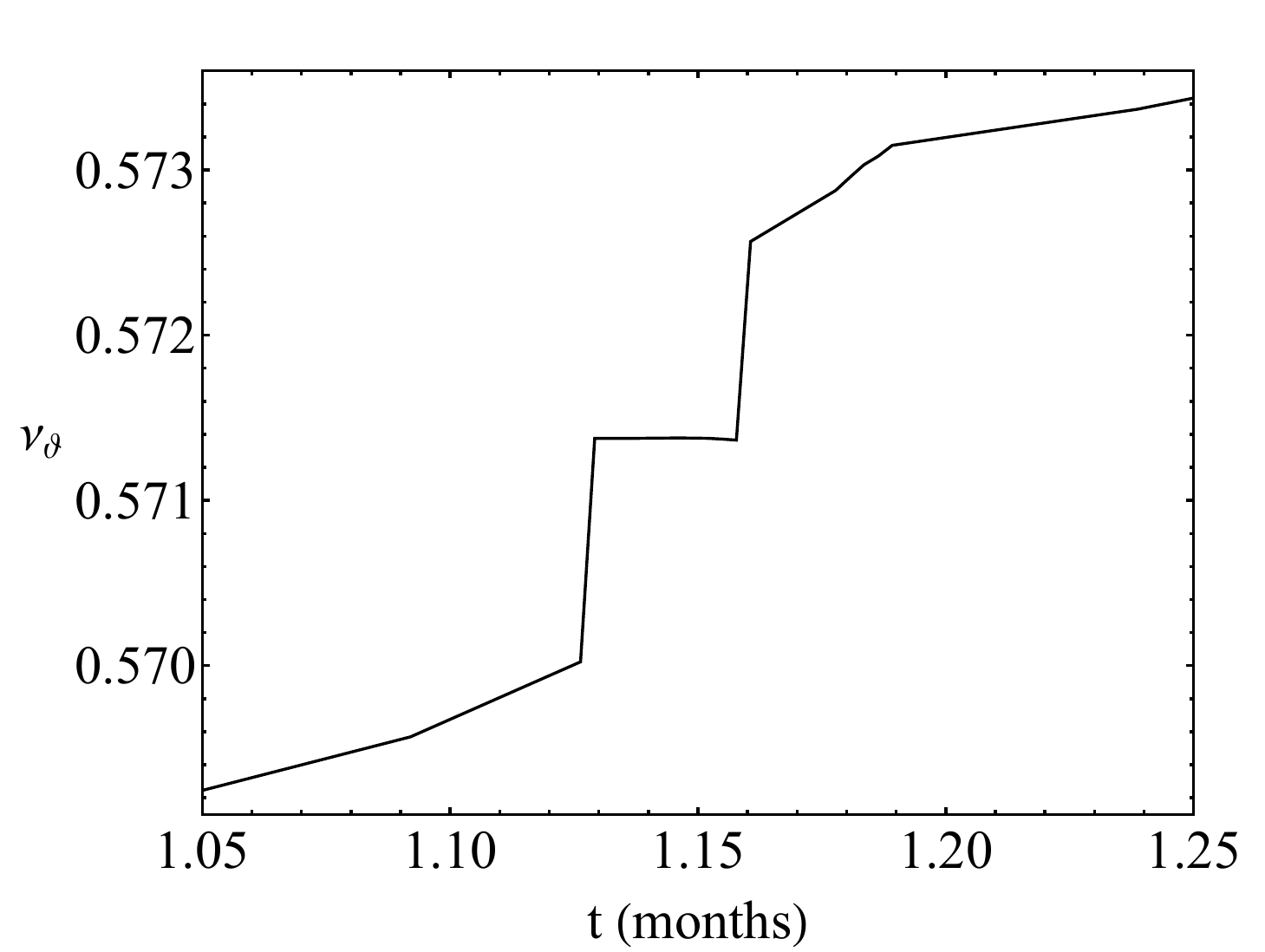}
\caption{Internal $4/7$ resonant island crossing for Case~3. To produce the inspiral we have updated the fluxes $250$ times. The secondary's initial parameters and conditions are $E(0)/m=0.95$, $L_z(0)/m=3M$, $r(0)=4.06025M$, $\dot{r}(0)=0$, $\theta(0)=\pi/2$ while the $\dot{\theta}(0)$ is defined from the constraint Eq.~\eqref{constraint_equation}. The EMRI spends $\sim100$ cycles in resonance.}\label{fig10}
\end{figure}

Figures~\ref{fig9} and~\ref{fig10} show typical, time-dependent, island crossings for Cases 1 and 3. In Fig.~\ref{fig9}, the non-zero initial velocity leads to the EMRI crossing the $2/3$ island in the exterior of Case~1 configuration. The plateau is finite and lasts for around 200 revolutions before it exits the island. The fact that the rotation number decreases with time shows that the orbital motion is outside of the star and tends towards circularization. On the other hand, in Fig.~\ref{fig10}, we show the dissipative rotation curve of the $4/7$ island crossing which resides inside the BS of Case~3. The fluxes have increased at this point significantly, therefore the EMRI evolves faster. Nevertheless, the secondary spends $\sim 100$ cycles in perfect resonance which is extraordinary for a subdominant resonance. Equivalently, the fact that these orbits are inside the star and both the eccentricity and rotation numbers increase, justifies why the dissipative rotation curve increases with time\footnote{It has been shown that, before plunge, Kerr EMRIs enter a short phase of increasing eccentricity~\cite{Gair:2005ih}. In our case, due to the absence of an event horizon, the eccentric EMRI can enter the star and the eccentricity continues increasing and leads to orbits as those found in~\cite{Grandclement:2014msa}. Since the eccentricity is directly proportional to the periapsis-apoapsis oscillation frequency $\omega_r$, the rotation number increases as well.}.
\begin{figure*}[t]
\includegraphics[scale=0.32]{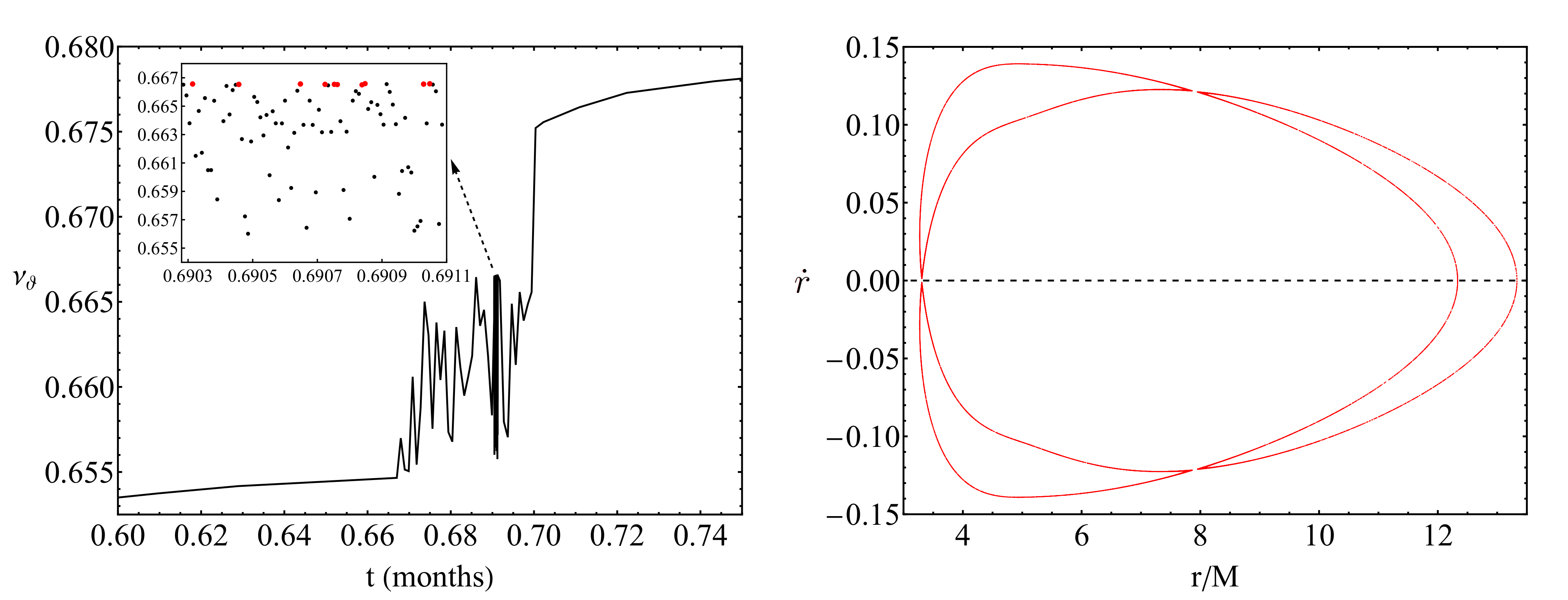}
\caption{Left: Interior $2/3$ resonant island crossing for Case~3. To produce the inspiral we have updated the fluxes $200$ times. The secondary's initial parameters and conditions are $E(0)/m=0.95$, $L_z(0)/m=3M$, $r(0)=6.5M$, $\dot{r}(0)=0.155$, $\theta(0)=\pi/2$, while $\dot{\theta}(0)$ is defined from the constraint Eq.~\eqref{constraint_equation}. The EMRI spends $\sim10$ cycles on and off the resonant island and $\sim 100$ revolutions in the vicinity the $2/3$ Birkhoff chain until the chaotic layer and island are crossed. Right: The produced KAM curves from the red dots shown in the inset of the left panel. All of them (namely, $10$ points) belong to the $2/3$ resonant island.}\label{fig11}
\end{figure*}

The final, and most interesting scenario examined for Case~3 is the very large plateau that arises when $\dot{r}(0)=0.155$, that begins from the exterior and ends in the interior of the BS. Even though one would expect a dissipative rotation curve with a plateau that lasts for thousands of cycles, the existence of a thick chaotic layer around the island induces a significant alteration to the rotation curve, which we find here for the first time. In Fig.~\ref{fig11} we present the behavior of the wide $2/3$ island. On the left, we show that instead of a plateau we encounter mostly the chaotic layer (regardless of the initial condition $r(0)$), where the rotation numbers are ill-defined. Nonetheless, there are points (marked in red in the zoom-in inset of the left panel in Fig.~\ref{fig11}) that actually enter the island. To make sure that these points do enter the resonance, we plot on the right the corresponding KAM curves of these points. As shown, they form islands with tips that do not touch, therefore the EMRI is not occupying the chaotic layer, unlike the rest of the black points in the inset of Fig.~\ref{fig11} that correspond to orbits residing in the chaotic layer. What seems to occur here is a strong chaotic shielding which does not allow the EMRI to spend continuous time intervals inside the island. Even so, we should expect interesting GW imprints for all cases discussed above and especially those from Case~3 which take place in the interior of the rotating BS. We present those effects in the next section.

We stress that our results are based on PN fluxes and neglect effects other than radiation-reaction within the BS, e.g. dynamical friction and accretion, which can nevertheless be added into the evolution.

\begin{figure*}[t]
\includegraphics[scale=0.315]{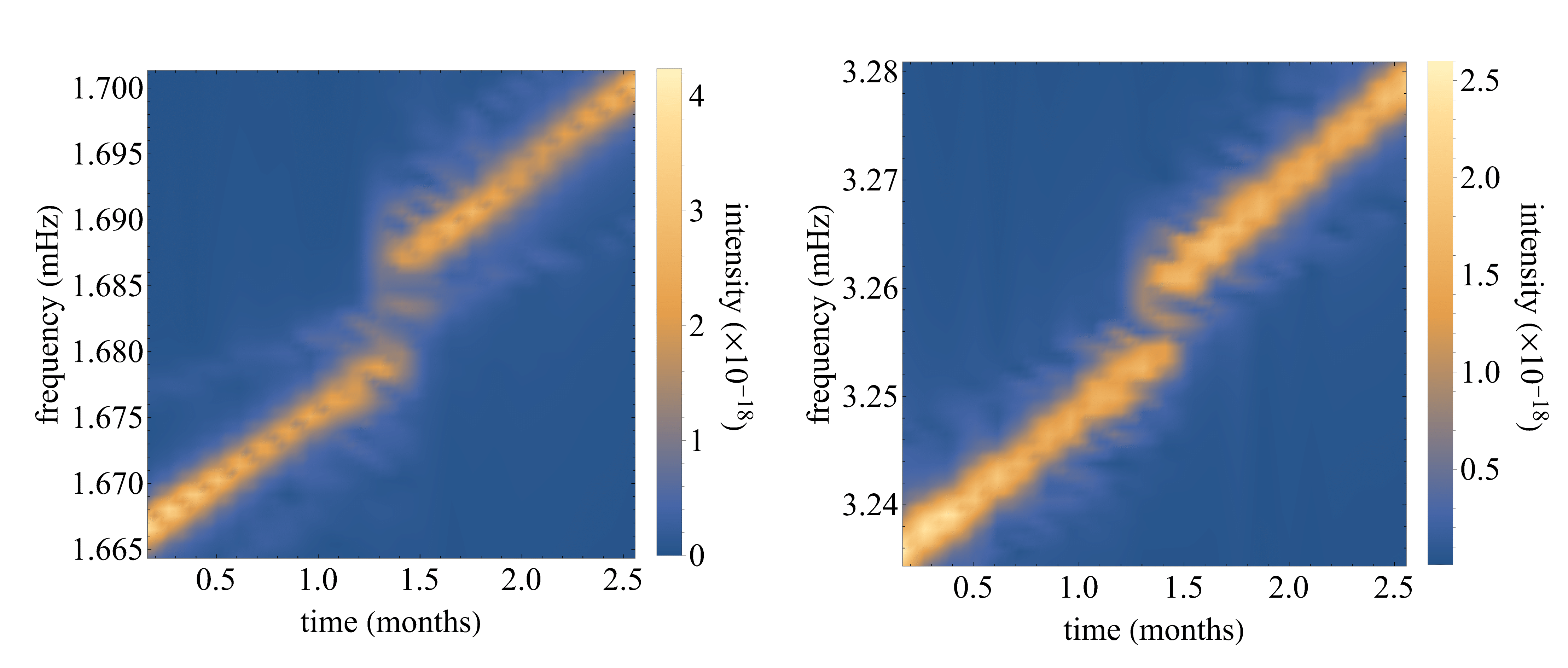}
\caption{Frequency evolution of an EMRI around rotating BS from Case~1, through the $2/3$ resonant island with parameters and initial conditions as in Fig.~\ref{fig9}. As a reference, the approximate GWs (fundamental frequency on the left panel and first harmonic on the right panel) are detected at luminosity distance $d=100\,\text{Mpc}$.}\label{fig12}
\end{figure*}
\begin{figure}[t]
\includegraphics[scale=0.36]{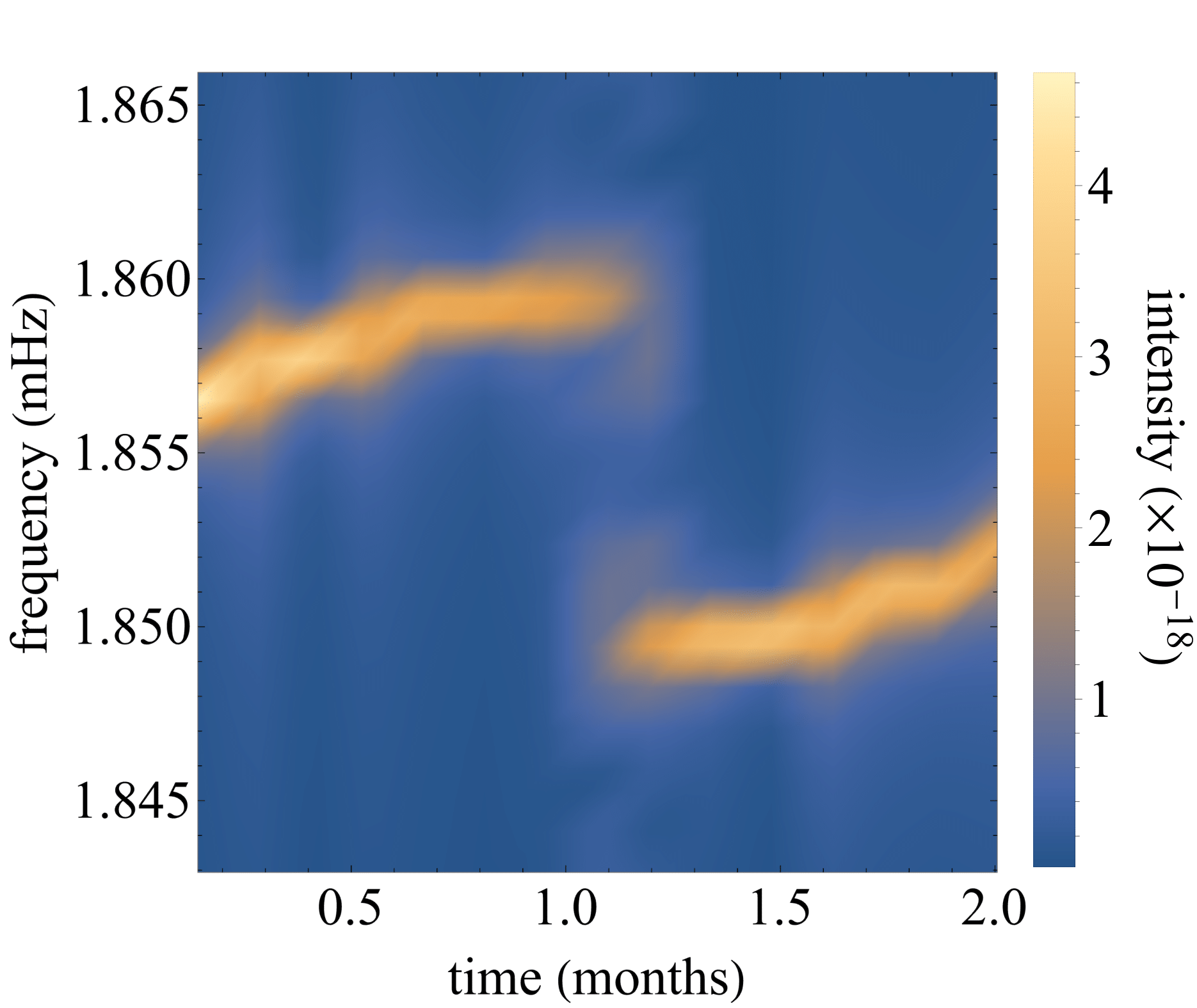}
\caption{Same as Fig.~\ref{fig12} but for the rotating BS configuration of Case~3, through the $4/7$ interior resonant island with parameters and initial conditions as in Fig.~\ref{fig10}.}\label{fig13}
\end{figure}
\begin{figure*}[t]
\includegraphics[scale=0.315]{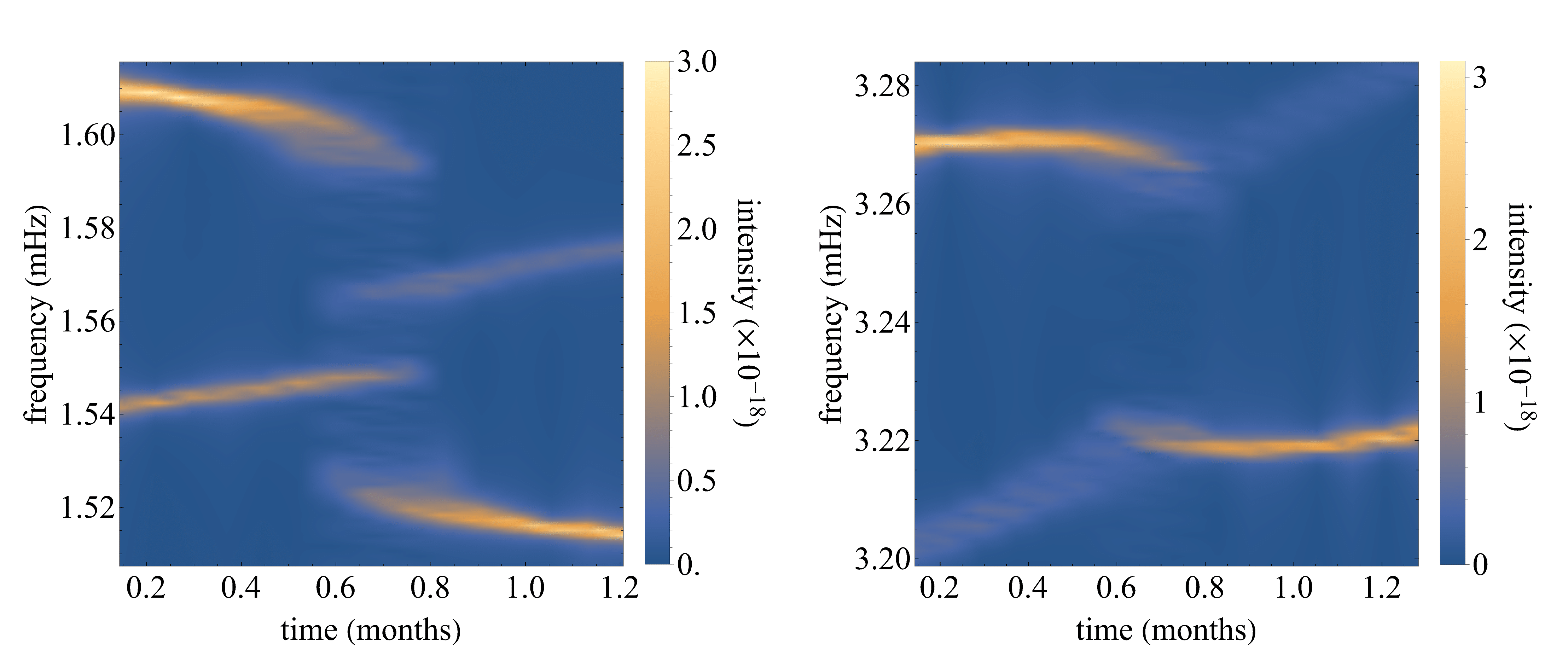}
\caption{Same as Fig.~\ref{fig12} but for the rotating BS configuration of Case~3, through the $2/3$ interior Birkhoff chain with parameters and initial conditions as in Fig.~\ref{fig11}.}\label{fig14}
\end{figure*}

\subsection{GW frequency evolution}

After obtaining the inspirals for Case~1 and 3, it is straightforward to use Eq.~\eqref{total_GW} in order to find the approximate GW emitted by such EMRIs. The time-domain waveforms are Fourier transformed to the frequency domain where the frequency evolution of the inspiral can be constructed. Here, we follow~\cite{Destounis:2021mqv,Destounis:2021rko,Destounis:2023gpw} and construct the spectrograms by performing consecutive short-time Fourier transforms, with appropriate window sizes and offsets in order to maximize the quality of the resulting figures.

In Fig.~\ref{fig12}, we present the most typical glitch waveform for Case~1, where the BS is rather compact and mimics a non-Kerr BH. Its fundamental and first-harmonic frequency evolution display the standard modulation that occurs when the EMRI crosses the $2/3$ island (that lies on the exterior of the rotating BS and lasts for 200 revolutions). This should lead to a significant dephasing with respect to an integrable EMRI evolution that crosses the same resonance. The resemblance with glitches found in~\cite{Destounis:2021mqv,Destounis:2021rko} for non-Kerr spacetimes is remarkable. Therefore, since we use the hybrid kludge scheme and the quadrupole formula, in this case it would be almost impossible to distinguish between non-Kerr BHs or rotating BS based only on these glitches. Fortunately, Case~3, the less compact configuration we have constructed, changes this picture completely. Figure~\ref{fig13} shows the frequency evolution of an EMRI through the internal $4/7$ island. The frequency evolution becomes nonlinear due to the acceleration of the fluxes and the EMRI frequencies after the glitch increase towards a chirp.

Finally, we examine the spectrogram of an EMRI crossing the internal $2/3$ resonant island that corresponds to the inspiral shown in Fig.~\ref{fig11}. The GW frequency modulation shown in Fig.~\ref{fig14} is the first one of its kind found in any non-integrable spacetime so far. We do not only observe similar nonlinear frequency evolution as a function of time for the strongest Fourier peaks (fundamental and first harmonic) as in Fig.~\ref{fig13}, but also linear-in-time subdominant frequency peaks that evolve and glitch in a similar manner. We have performed the same inspiral for two different initial conditions $r(0)$, one that lies outside and one that lies inside the range of sustained resonances (if linear fluxes are assumed) and the resulting spectrogram is qualitatively the same as that shown in Fig.~\ref{fig14}. This feature is significantly different from all the other glitches found in previous non-integrable spacetimes, and is associated to the fact that the plateau in the dissipative rotation curve is replaced with an observable chaotic layer with glimpses of island occupancy. 

When an orbit is fully chaotic and resides in a chaotic sea \cite{Contopoulos_book}, then we expect rotation curves and spectrograms to be rendered useless since no well-defined rotation numbers and discrete GW frequencies exist~\cite{Kiuchi:2004bv,Gair:2007kr}. On the other hand, when an orbit is close to a slightly chaotic region but otherwise not a fully chaotic part of phase space, such as the chaotic layer that shields the interior $2/3$ island of Case~3, the waveform and its frequency content resemble a lot those of a regular orbit, with discrete Fourier peaks and indistinguishable effects of chaos in the time-domain waveform. The only key difference is that the discrete Fourier peaks are not comprised by single harmonics (for geodesics) but rather harmonics that are broken down to subpeaks. The spectrum, nevertheless, remains discrete. This has been shown in~\cite{Kiuchi:2004bv} for a secondary particle with spin orbiting around a Schwarzschild BH (see Figs. 4, 10, 11 and 13 therein), although in that case this behavior disappears as the secondary spin is treated perturbatively, as also requested for consistency within the perturbative expansion in the mass ratio (see, e.g.,~\cite{Piovano:2020zin}). It is therefore interesting that we recover the same feature, that is a slightly chaotic region that embeds the island, but otherwise normal generic orbits around this region, discrete Fourier spectra, and harmonics with multiple peaks, but in rotating BS spacetimes (where motion is not even integrable in a perturbative sense). This is the reason why Fig.~\ref{fig14} has such a distinctive and peculiar nature, which can be explained by the effects discussed in~\cite{Kiuchi:2004bv}. If we consider the effects on the spectrogram and the dissipative evolution through the Birkhoff chain, the EMRIs spends around 100 cycles to cross it, and 10 cycles in perfect resonance. Yet, a Birkhoff chain is composed of both stable and unstable periodic points, and the chaotic layer helps to amplify the frequency glitch in a significant way, with frequency jumps as large as $\sim 0.07$ mHz for the most dominant fundamental Fourier peak evolution, which is almost one order of magnitude larger than any other frequency glitches presented here, and also higher than those found in~\cite{Destounis:2021mqv,Destounis:2021rko} for various deformation parameters that were either exaggerated or chosen arbitrarily. Our case presents a well motivated model of ECOs that has a precise formation scenario, thus the results presented here can be used in the future as a testbed to understand if weak or strong (or no) chaos exists in GW observations.

\section{Conclusions}

The detection of EMRIs by LISA~\cite{LISA:2017pwj} and other spaceborne detectors~\cite{Ruan:2018tsw,Ruan:2020smc,TianQin:2015yph} will provide for the first time an extremely accurate mapping of the primary's spacetime geometry, due to the large mass hierarchy that creates the conditions for long-lasting inspirals with continuously observable GW emission. The associated signal will present very rich phenomenology~\cite{Amaro-Seoane:2022rxf} like the appearance of resonances~\cite{Apostolatos:2009vu,Contopoulos:2011dz,Lukes-Gerakopoulos:2012qpc,Lukes-Gerakopoulos:2021ybx} and, remarkably, carries information about spacetime symmetries~\cite{Destounis:2020kss,Lukes-Gerakopoulos:2010ipp}.

The observation of these systems, will also help to decide if BHs are the only compact objects in the universe or if other exotic configurations exist in nature. The different multipolar structure, relative to Kerr BHs, can lead to fascinating phenomena around resonances such as frequency modulation and GW glitches \cite{Destounis:2021mqv}, in particular for those spacetimes that lack a Carter constant (or any other higher-rank Killing tensor) that would guarantee the integrability of geodesics. So far, only bumpy and non-Kerr solution of GR --~that are plagued with pathologies~\cite{Destounis:2021rko,Destounis:2023gpw} and do not come from a well motivated, first-principle theory~-- have been analyzed in order to inspect their different phenomenology.

Here, we considered for the first time one of the most well-behaved and simple compact object that may as well exist in our universe, namely rotating BSs. These exotic objects which lack an event horizon or singularities, have a clear formation mechanism and can serve as prototypical BH mimickers. They also constitute compelling dark matter candidates, and could have formed in the early Universe, e.g. as remnants of inflation.  

While the majority of analyses for BSs have been performed for equatorial orbits, here we took a different turn on examining supermassive self-interacting rotating BSs in the context of EMRIs for generic orbits (without constraints on the orbital plane or eccentricity). In an attempt to study geodesics that are peculiar enough to be distinguishable from those around Kerr BHs, we evolved geodesics around and inside supermassive rotating configurations and studied their GW signatures. We have built numerically three of such configurations with representative values of the spin and compactness.

Our geodesic analysis reveals the existence of resonant islands (with finite width) and Birkhoff chains in the phase space of orbits, which signals that rotating BSs are non-integrable. We also found that compact rotating BSs behave similarly to non-Kerr BHs up to the point where the geodesic enters the star smoothly and the godesic structure becomes qualitatively different. Decreasing the compactness makes things much more different between rotating BSs and non-Kerr BHs. First, the geodesic phase space hosts many orbits that reside in the star's interior, where the rotation curve exhibits a change in monotonicity and begins to increase due to the increment of eccentricity~\cite{Grandclement:2014msa}. The latter effect leads to a newly-found phenomenon, that is the existence of degenerate islands of stability, in the interior and the exterior of the star, with the same rotation number. We have also found regions in the phase space of geodesics where there exist not only a plethora of resonant islands, but also a thin chaotic layer surrounding the most dominant $2/3$ island of stability. This is, to our knowledge, the first time that a full Birkhoff chain appears in a general relativistic setup, which has important implications in GWs emitted by EMRIs that cross this region. With appropriate initial conditions, we find that the $2/3$ island encounters a plateau in the rotation curve, where the geodesic enters the island in the exterior and exits in the interior of the star; a fascinating outcome of the fact that the compactness of the BS of Case~3 (albeit as large as a typical neutron star) is relatively smaller than in the other cases.

To achieve some initial estimates of the elementary structure of approximate waveforms from such rotating BSs, and especially for generic inspirals that cross transient resonances, we use the quadrupole approximation to model GWs and evolve the inspiral, with augmented PN fluxes to account for the modified multipolar structure of our configurations~\cite{Vaglio:2022flq,Vaglio:2023lrd}. 
Despite these approximations, the phenomenology of the inspiral and GW signal is expected to be robust.

At first sight with a linear approximation of the fluxes, we stumble upon inspirals that enter the $2/3$ island, in a generous range of initial conditions $r(0)$, and never exit the resonance, as also found in~\cite{Lukes-Gerakopoulos:2021ybx} but in a much smaller radial domain. However, when we consistently update fluxes during the inspiral (though in a discretized way after some amount of revolutions), higher order terms seem to destroy such sustained resonances. Although unlikely, we cannot unequivocally exclude the existence of other regions of phase space where such resonances occur even when updates in the fluxes are introduced. We sketch the dissipative rotation curves by using time instants of each EMRI as initial condition to a geodesic evolution and find that external resonant islands are indistinguishable with those occurring in non-Kerr objects~\cite{Destounis:2021mqv,Destounis:2021rko,Destounis:2023gpw}. 
The plateaus that characterize islands situated in the interior of the star are approached from below in the rotation curves, which then reprise a monotonically increasing behavior. This corresponds to a reversed time and radial dependence with respect to the rotation curves in the neighbourhood of external islands, related to the non-zero energy-density distribution inside the star. Such monotonicity reversal also pairs with the different evolution of the eccentricity. Indeed, inside the star, eventually a point is reached where the circularization of orbits have lapsed and the eccentricity increases with decreasing radius, in contrast to what happens in the exterior. The most prominent island of all found to date, the one for Case~3 that has a record width of $2.6M$ (in quasi-isotropic coordinates), is even more complicated due to chaotic shielding. The EMRI does not enter and stay in the island but rather oscillates between on and off island states. It manages to do so at least 10 times during which the evolution drives the orbit around the chaotic layer till the secondary eventually exits the particular part of the Birkhoff chain. 

Overall, the time spent in these cases spans from $200$ cycles in external resonances, where the evolution is adiabatically slow, to around $100$ cycles in the interior of the star, when the Birkhoff chain is taken into account as a part of the whole chaotic KAM curve (and not just the island with the periodic stable point at its center).

When the EMRI crosses a resonant island, all the above effects are imprinted onto the EMRI waveform either in a typical way, analogous to other non-Kerr systems, or as novel signatures that designate the existence of a supermassive BS primary. GW signals from the exterior islands are qualitatively indistinguishable from those in non-Kerr EMRIs; a result which is expected from the close similarity in the rotation curves discussed above. The case of internal motion is instead quite different. Even subdominant resonant islands introduce significant nonlinear frequency modulation because the orbits is close to exit the CZV (see e.g. Fig.~\ref{fig13}). The widest resonant island of Case~3 has a very special behavior imprinted in the GW when the EMRI crosses it. The proximity of the orbits to the otherwise mild (but still observable) chaotic layer leads to an effect in the GW Fourier peaks first found in~\cite{Kiuchi:2004bv}. Moderate chaotic layers do not produce continuous GW spectra, as fully chaotic orbits do, but rather discrete sets of harmonics that are broken down to sub-harmonics. This phenomenon is evident in Fig.~\ref{fig14} where two simultaneous glitches occur under to the evolution of each sub-frequency due to mild chaos. The underlying reason is, again, the fact that the EMRI is close to crossing a thin chaotic region. This turns each single harmonic of the frequency content of the approximated GW signal into multiple sub-peaks, that arise in the spectrogram in the particular fashion shown in Fig.~\ref{fig14}. The most prominent (the brightest) one, which leads to a frequency glitch of ${\cal O}(0.1)\,{\rm mHz}$, renders all exterior glitches subdominant and should affect significantly the orbital evolution and eventual parameter estimation. Yet, we stress that our results are qualitative and should be extended with more accurate inspiral and waveform modeling.

In this respect it would be very interesting to perform proper perturbation theory around a spinning BS, although nonseparability of the equations makes the flux computation quite challenging compared to the Teukolsky case for a Kerr BH. Nevertheless our results provide solid evidence that chaos does exist in EMRIs around rotating BSs, and the fact that this is a quite compelling model for ECOs should drive us to further understand these objects, as well as their effects in geodesics and EMRI dynamics.

Along these lines it would be relevant to include environmental effects such as dynamical friction and accretion~\cite{Macedo:2013jja,Cardoso:2021wlq,Cardoso:2022whc,Destounis:2022obl} and assess their impact on the chaotic motion and EMRI signal. Finally, although we focused on a specific model of BS with large quartic interactions, we expect to find the same qualitative features in other spinning BS models, spinning Proca stars, and in fact in all spinning ECOs where geodesic motion is most likely nonintegrable. In particular, we expect the same (and perhaps even more prominent) signatures of chaotic motion for EMRIs in generic orbits around BH microstates predicted in the fuzzball scenario (see~\cite{Bena:2022ldq} for a recent review), due to their rich multipolar structure and breaking of the axial and equatorial symmetry~\cite{Bena:2020see,Bianchi:2020bxa,Bena:2020uup,Bianchi:2020miz,Fransen:2022jtw}.

\begin{acknowledgements}
The authors are indebted to Georgios Lukes-Gerakopoulos and Kostas D. Kokkotas for insightful discussions. We acknowledge financial support provided under the European Union's H2020 ERC, Starting Grant agreement no.~DarkGRA--757480 and support under the MIUR PRIN (Grant 2020KR4KN2 “String Theory as a bridge between Gauge Theories and Quantum Gravity”) and FARE (GW-NEXT, CUP: B84I20000100001, 2020KR4KN2) programmes. K.D. is supported by the DAAD program for the ``promotion of the exchange and scientific cooperation between Greece and Germany IKYDAAD 2022" (57628320).
\end{acknowledgements}

\bibliography{References}
	
\end{document}